%% file: main.tex
\documentclass{elsarticle}

\usepackage[colorlinks=true,%
             linkcolor=blue,%
            citecolor=blue,%
            urlcolor=blue]{hyperref}   

\usepackage{xspace}
\usepackage{epsfig}
\usepackage{latexsym,amssymb,stmaryrd}
\usepackage{fancybox,epic}
\usepackage{amsmath}
\usepackage{colortbl}
\usepackage{color} 
\usepackage{graphics}
\usepackage{epsfig}
\usepackage{graphicx}
\usepackage[draft]{fixme}
\usepackage{tikz}
\usepackage{wrapfig}

\usepackage{listings}
\makeatletter
\def\ps@pprintTitle{%
  \let\@oddhead\@empty
  \let\@evenhead\@empty
  \let\@oddfoot\@empty
  \let\@evenfoot\@oddfoot
}
\makeatother

\newtheorem{theorem}{Theorem}[section]

\newtheorem{example}[theorem]{Example}
\newtheorem{definition}[theorem]{Definition}
\newtheorem{lemma}[theorem]{Lemma}

\newtheorem{remark}[theorem]{Remark}

\usepackage{fancyhdr} 
\newcommand{\cI}[0]{{\cal I}}

\newcommand{\cK}[0]{{\cal K}}
\newcommand{\cL}[0]{{\cal L}}

\newcommand{\cO}[0]{{\cal O}}

\newcommand{\cR}[0]{{\cal R}}

\newcommand{\cV}[0]{{\cal V}}

\newcommand{\cX}[0]{{\cal X}}
\newcommand{\cY}[0]{{\cal Y}}

\def\tuple#1{\langle{#1}\rangle}

\newcommand{\WO}[2]{\exists #2.#1}
\newcommand{\SHARE}[2]{\tuple{#1{\bullet}#2}}

\newcommand{\ALIAS}[2]{\tuple{#1{\cdot}#2}}

\newcommand{\REACHES}[2]{#1{\rightsquigarrow}#2}

\newcommand{\CYCLE}[1]{{\circlearrowleft}^{#1}}

\DeclareMathOperator{\interp}{\iota}
\DeclareMathOperator{\dom}{dom}
\DeclareMathOperator{\codom}{rng}

\newcommand{\body}[1]{\ensuremath{{#1}^b}}
\newcommand{\inp}[1]{\ensuremath{{#1}^i}}

\newcommand{\locals}[1]{\ensuremath{{#1}^l}}
\newcommand{\scope}[1]{\ensuremath{{#1}^s}}
\newcommand{\nil}{\mbox{\lstinline!null!}\xspace}
\newcommand{\xx}{\mbox{\lstinline!x!}\xspace}
\newcommand{\yy}{\mbox{\lstinline!y!}\xspace}
\newcommand{\zz}{\mbox{\lstinline!z!}\xspace}

\newcommand{\nn}{\mbox{\lstinline!n!}\xspace}
\newcommand{\pp}{\mbox{\lstinline!p!}\xspace}
\newcommand{\cc}{\mbox{\lstinline!c!}\xspace}
\newcommand{\ii}{\mbox{\lstinline!i!}\xspace}
\newcommand{\codepos}{\mbox{\lstinline!pos!}\xspace}
\newcommand{\codethis}{\mbox{\lstinline!this!}\xspace}
\newcommand{\integer}{\mbox{\lstinline!int!}\xspace}
\newcommand{\newk}[1]{\ensuremath{\mbox{\lstinline!new!}~#1}}
\newcommand{\assign}{\mbox{\lstinline!:=!}}
\newcommand{\ifte}{\mbox{\lstinline!if!}\xspace}
\newcommand{\iftethen}{\mbox{\lstinline!then!}\xspace}
\newcommand{\ifteelse}{\mbox{\lstinline!else!}\xspace}
\newcommand{\while}{\mbox{\lstinline!while!}\xspace}
\newcommand{\whilebody}{\mbox{\lstinline!do!}\xspace}
\newcommand{\return}{\mbox{\lstinline!return!}\xspace}
\newcommand{\this}{\ensuremath{{\mathord{\mathit{this}}}}}
\newcommand{\out}{\ensuremath{{\mathord{\mathit{out}}}}}

\newcommand{\variables}{\ensuremath{\cX}}
\newcommand{\identifiers}{\ensuremath{\cX}}
\newcommand{\classes}{\ensuremath{\cK}}
\newcommand{\class}{\ensuremath{\kappa}}
\newcommand{\objects}{\ensuremath{\cO}}

\newcommand{\subclasseq}{\preceq}

\newcommand{\methodsig}[1]{\ensuremath{\mathsf{#1}}}

\newcommand{\values}{\ensuremath{\cV}}
\newcommand{\locations}{\ensuremath{\cL}}
\newcommand{\heap}{\ensuremath{\mu}}
\newcommand{\objtag}[1]{\ensuremath{{#1}.\mathsf{tag}}}
\newcommand{\objframe}[1]{\ensuremath{{#1}.\mathsf{frm}}}
\newcommand{\newobj}[1]{\ensuremath{\mathit{newobj}(#1)}}
\newcommand{\newloc}[1]{\ensuremath{\mathit{newloc}(#1)}}

\newcommand{\statesym}{\ensuremath{\sigma}}
\newcommand{\state}[2]{\ensuremath{\tuple{#1,#2}}}
\newcommand{\states}[1]{\ensuremath{\Sigma_{#1}}}
\newcommand{\frm}{\ensuremath{\phi}}
\newcommand{\statef}[1]{\ensuremath{\hat{#1}}}
\newcommand{\statem}[1]{\ensuremath{\check{#1}}}

\newcommand{\den}{\ensuremath{\delta}}
\newcommand{\denset}[2]{\ensuremath{\Delta(#1{,}#2)}}
\newcommand{\einter}[3]{\ensuremath{{E_{#1} ^{#2}}{\left\llbracket{#3}\right\rrbracket}}}
\newcommand{\cinter}[3]{\ensuremath{{C_{#1}^{#2}}{\left\llbracket{#3}\right\rrbracket}}}

\newcommand{\seq}{\ensuremath{\mapsto}}
\newcommand{\lookup}{\ensuremath{\mathit{lkp}}}

\newcommand{\tp}[1]{\ensuremath{T_{#1}}}

\newcommand{\project}[2]{\ensuremath{\exists{#2}.{#1}}}
\newcommand{\extend}[2]{\ensuremath{extend({#1},{#2})}}
\newcommand{\res}{\ensuremath{\rho}}

\newcommand{\abstp}[1]{\ensuremath{\mathcal{T}_{#1}}}

\newcommand{\typenv}{\ensuremath{\tau}}
\newcommand{\interpretations}{\ensuremath{\Gamma}}

\newcommand{\reachable}[2]{\ensuremath{R(#1,#2)}}
\newcommand{\reachablei}[3]{\ensuremath{R^{#1}(#2,#3)}}

\newcommand{\ereachable}[2]{\ensuremath{R^{\varepsilon}(#1,#2)}}

\newcommand{\ASEMANTICS}[3]{\ensuremath{\mathcal{C}_{#2}^{#1}}{\left\llbracket{#3}\right\rrbracket}}
\newcommand{\EXPASEMANTICS}[3]{\ensuremath{\mathcal{E}_{#2}^{#1}}{\left\llbracket{#3}\right\rrbracket}}

\newcommand{\condom}{\ensuremath{\cI_\flat^\typenv}}

\newcommand{\domr}{\ensuremath{\cI_r^\typenv}}
\newcommand{\gammar}{\ensuremath{\gamma_{\mathit r}^\typenv}}
\newcommand{\alphar}{\ensuremath{\alpha_{\mathit r}^\typenv}}

\newcommand{\domc}{\ensuremath{\cI_c^\typenv}}
\newcommand{\gammac}{\ensuremath{\gamma_{\mathit c}^\typenv}}
\newcommand{\alphac}{\ensuremath{\alpha_{\mathit c}^\typenv}}

\newcommand{\domrc}[1]{\ensuremath{\cI_{rc}^{#1}}}
\newcommand{\gammarc}{\ensuremath{\gamma_{\mathit rc}^\typenv}}
\newcommand{\alpharc}{\ensuremath{\alpha_{\mathit rc}^\typenv}}

\newcommand{\concelem}{\ensuremath{I_\flat}}
\newcommand{\abselemr}{\ensuremath{I_r}}
\newcommand{\abselemc}{\ensuremath{I_c}}
\newcommand{\abselemrc}{\ensuremath{I_{rc}}}

\newcommand{\reachset}[1]{\ensuremath{\cR^{#1}}}
\newcommand{\cycset}[1]{\ensuremath{\cY^{#1}}}

\newcommand{\absinterp}{\ensuremath{\zeta}}
\newcommand{\absden}{\ensuremath{\xi}}
\newcommand{\absdenset}[2]{\ensuremath{\Xi({#1},{#2})}}
\newcommand{\absinterpretations}{\ensuremath{\Psi}}

\newcommand{\shden}[1]{\ensuremath{\mathsf{SP}_{#1}}}
\newcommand{\PURE}[1]{\ensuremath{\dot{#1}}}

\begin{document}

\title{Reachability-based Acyclicity Analysis by Abstract Interpretation}


\author{SAMIR GENAIM\\
  Complutense University of Madrid\\
  DAMIANO ZANARDINI \\
  Technical University of Madrid
}






\setcounter{page}{1}

\input abstract

\begin{keyword}
  Abstract Interpretation; Acyclicity Analysis; Termination Analysis;
  Object-Oriented Programming; Heap Manipulation
\end{keyword}

\maketitle

\input intro

\input running-example
\input oo-language-short

\input domain
\input analysis-short
\input experiments

\input conclusions

\paragraph{Acknowledgments}

This work was funded in part by the Information \& Communication
Technologies program of the European Commission, Future and Emerging
Technologies (FET), under the ICT-231620 {\em HATS} project, by the
Spanish Ministry of Science and Innovation (MICINN) under the
TIN2008-05624, TIN2012-38137 and PRI-AIBDE-2011-0900 projects, by
UCM-BSCH-GR35/10-A-910502 grant and by the Madrid Regional Government
under the S2009TIC-1465 \emph{PROMETIDOS-CM} project.

\newpage
\bibliographystyle{abbrv}


\newpage
\appendix
\input proofs

\end{document}

%% file: abstract.tex

\begin{abstract}
  In programming languages with dynamic use of memory, such as Java,
  knowing that a reference variable \xx points to an acyclic data
  structure is valuable for the analysis of \emph{termination} and
  \emph{resource usage} (e.g., execution time or memory consumption).
  For instance, this information guarantees that the \emph{depth} of
  the data structure to which \xx points is greater than the depth of
  the data structure pointed to by \xx.$f$ for any field $f$ of \xx.
  This, in turn, allows bounding the number of iterations of a loop
  which traverses the structure by its depth, which is essential in
  order to prove the termination or infer the resource usage of the
  loop.
  The present paper provides an Abstract-Interpretation-based
  formalization of a static analysis for inferring acyclicity, which
  works on the \emph{reduced product} of two abstract domains:
  \emph{reachability}, which models the property that the location
  pointed to by a variable $w$ can be reached by dereferencing another
  variable $v$ (in this case, $v$ is said to reach $w$); and
  \emph{cyclicity}, modeling the property that $v$ can point to a
  cyclic data structure.  The analysis is proven to be \emph{sound}
  and \emph{optimal} with respect to the chosen abstraction.
  %
  %
\end{abstract}

%% file: intro.tex

\section{Introduction}
\label{sec:introduction}

Programming languages with dynamic memory allocation, such as Java,
allow creating and manipulating cyclic data structures. The presence
of cyclic data structures in the program memory (the \emph{heap}) is a
challenging issue in the context of termination
analysis~\cite{DBLP:conf/cav/BerdineCDO06,DBLP:conf/pldi/CookPR06,AlbertACGPZ08,SpotoMP2010},
resource usage analysis~\cite{Wegbreit75,caslog,AlbertAGPZ12}, garbage
collection~\cite{JonesLins1996}, etc.
As an example, consider the loop
``\lstinline!while (x$!$=null) do x:=x.next;!'': if \xx points to an
acyclic data structure before the loop, then the depth of the data
structure to which \xx points strictly decreases after each iteration;
therefore, the number of iterations is bounded by the initial depth of
(the structure pointed to by) \xx.  On the other hand, the possibility
that \xx points to a cyclic data structure forbids, in general,
proving that the loop terminates.

Automatic inference of such information is typically done by (1)
\emph{abstracting} the loop to a numeric loop ``${\it while}(x)
\leftarrow \{x{>}0,x{>}x'\},{\it while}(x')$''; and (2) bounding the
number of iterations of the numeric loop.
The numeric loop means that, if the loop entry is reached with \xx
pointing to a data structure with depth $x>0$, then it will eventually
be reached again with \xx pointing to a structure with depth $x'<x$.
The key point is that ``\lstinline!x$!$=null!'' is abstracted to
$x>0$, meaning that the depth of a non-null variable cannot be 0;
moreover, abstracting ``\lstinline!x:=x.next!'' to $x>x'$ means that
the depth decreases when accessing fields.  While the former is valid
for any structure, the latter holds only if \xx is acyclic.
Therefore, acyclicity information is essential in order to apply such
abstractions.

In mainstream programming languages with dynamic memory manipulation,
data structures can only be modified by means of \emph{field updates}.
If, before \lstinline!x.f:=y!, \xx and \yy are guaranteed to point to
disjoint parts of the heap, then there is no possibility to create a
cycle.  On the other hand, if they are not disjoint, i.e.,
\emph{share} a common part of the heap, then a cyclic structure might
be created.
This simple mechanism has been used in previous
work~\cite{RossignoliS06} in order to declare \xx and \yy, among
others, as (possibly) cyclic whenever they share before the update.
In the following, we refer to this approach as the
\emph{sharing}-based approach to acyclicity analysis.

The sharing-based approach to acyclicity is simple and efficient,
however, there can be an important loss of precision in typical
programming patterns.  E.g., consider
``\lstinline!y:=x.next.next; x.next:=y;!'', which typically removes an
element from a linked list, and let \xx be initially acyclic.  After
the first command, \xx and \yy clearly share, so that they should be
declared as finally cyclic, even if, clearly, they are not.
When considering \lstinline!x.f:=y!, the precision of the acyclicity
information can be improved if it is possible to know \emph{how} \xx
and \yy share.  To this end, there are four possible scenarios: (1)
\xx and \yy alias; (2) \xx reaches \yy; (3) \yy reaches \xx; (4) they
both reach a common location.  The field update \lstinline!x.f:=y!
might create a cycle only in cases (1) and (3).
An acyclicity analysis based on similar observations has been
considered before in the context of C
programs~\cite{DBLP:conf/popl/GhiyaH96}, where the analysis has been
presented as a \emph{data-flow} analysis, however, no formal
justification for its correctness has been provided.
In what follows, we refer to this approach as the
\emph{reachability}-based approach to acyclicity analysis.

\subsection{Contributions}

The main contribution of this paper is essentially theoretical.  In
particular, the paper formalizes an existing reachability-based
acyclicity analysis~\cite{DBLP:conf/popl/GhiyaH96} within the
framework of Abstract Interpretation~\cite{Cousot77}, and proves its
soundness and optimality:

\begin{enumerate}
\item We define an \emph{abstract domain} $\domrc{\typenv}$, which
  captures the reachability information about program variables (i.e.,
  whether there can be a path in the heap from the location $\ell_v$
  bound to some variable $v$ to the location $\ell_w$ bound to some
  $w$), and the acyclicity of data structures (i.e., whether there can
  be a cyclic path starting from the location bound to some variable).
\item A provably sound and optimal {\it abstract semantics}
  $\ASEMANTICS{\typenv}{\absinterp}{\_}(\_)$ of a simple
  object-oriented language is developed, which works on
  $\domrc{\typenv}$ and can often guarantee the acyclicity of {\it
    Directed Acyclic Graphs} (DAGs), which most likely will be
  considered as cyclic if only sharing, not reachability, is taken
  into account.  With respect to the original analysis, the definition
  of the abstract semantics involves additional effort like dealing
  with specific features of object-oriented languages, and discussing
  some technical improvements.
\end{enumerate}

\noindent
As a proof of concept, the abstract semantics has been also
implemented in the COSTA~\cite{AlbertAGPZ07d} COSt and Termination
Analyzer as a component whose result is an essential information for
proving the termination or inferring the resource usage of programs
written in Java bytecode.  Focusing on full Java bytecode, the
implementation has also to deal with advanced features of the language
like exceptions and static fields.

The present paper is based on preliminary work by the same authors
which was published as a short workshop version \cite{GenaimZ10} and
as a technical report \cite{GenaimZ10tr}.

\input related

\subsection{Organization}

The rest of the paper is organized as follows:
Section~\ref{sec:runningExample} presents an example of
reachability-based acyclicity analysis.  Section.~\ref{sec:language}
defines the syntax and semantics of a simple Java-like language.
Section~\ref{sec:abstractDomain} introduces the abstract domains for
reachability and cyclicity, and their reduced product, and
Section~\ref{sec:abstractSemantics} defines the abstract semantics,
and proves some important properties.  Finally,
Section~\ref{sec:conclusions} concludes the paper.
Proofs of the technical results are available in~\ref{sec:proofs}.

%% file: related.tex

\subsection{Related work}
\label{sec:relatedWork}

A reachability-based acyclicity analysis for C programs was developed
in~\cite{DBLP:conf/popl/GhiyaH96}; however, that analysis was
presented as a \emph{data-flow} analysis, and it did not include any
formal justification of its correctness.
Our paper provides a formalization of a similar analysis in terms of
Abstract Interpretation, and includes soundness proofs.  Note
that~\cite{DBLP:conf/popl/GhiyaH96} uses the terms ``direction'' and
``interference'', respectively, for reachability and sharing.


As far as Abstract-Interpretation-formalized {\it cyclicity analyses}
are concerned, the one by Rossignoli and Spoto \cite{RossignoliS06} is
the most related work.  This analysis is only based on sharing (not on
reachability), and, as discussed in the paper, is less precise than
the reachability-based approach.

The work on {\it Shape Analysis} \cite{WilhelmSR00-long} is related
because it reasons about heap-manipulating programs in order to prove
program properties.  In most cases, {\it safety} properties are dealt
with \cite{BardinFN04,SagivRW02,RinetzkyBRSW05}.  On the other hand,
termination is a {\it liveness} property, and is typically the final
property to be proven when analyzing acyclicity.  Therefore, work on
liveness properties will be considered more deeply.  Most papers
\cite{Reynolds02,BalabanPZ05,DBLP:conf/cav/BerdineCDO06,DBLP:conf/pldi/CookPR06,BrotherstonBC08-short}
use techniques based on like {\it Model Checking}
\cite{Muller-OlmSS99}, {\it Predicate Abstraction} \cite{GrafS97},
     {\it Separation Logic} \cite{Reynolds02} or {\it Cyclic proofs}
     \cite{BrotherstonBC08-short} to prove properties for programs
     which work on
{\it single-linked} heaps.  This means that only one heap cell is
directly reachable from another one, which is basically the same as
having, in an object-oriented language, only one class with one field.
This somehow restricts the structure of the heap and, in some cases,
allows obtaining more precise results.  On the contrary, the present
paper deals with a technique which does not rely on such an
assumption: as the language is object-oriented, every object can have
multiple fields.
Other works \cite{BalabanPZ07} deal with {\it single-parent} heaps,
which are multi-linked but sharing-free; needless to say, the present
paper handles heap structures where sharing is more than a
possibility.  There also exist other works \cite{GotsmanBC06} based on
Separation Logic which efficiently prove program properties and deal
with cyclic structures, but are specialized to a limited set of data
structures like single-linked lists, double-linked lists or trees.
Also, in most of these works, the heap size is bounded by some
constant, which is also a minor limitation.  On the contrary, the
present paper deals with data structures which can have practically
any shape, and tries to infer information about the shape on its own.
It is convenient to point out that the acyclicity analysis under
discussion does not focus on directly proving liveness properties;
instead, it is supposed to provide useful information to a
cost\footnote{This analysis is actually implemented in COSTA, which
  handles both cost and termination of the Java bytecode programming
  language} or termination analyzer which will perform the task.

%% file: running-example.tex

\section{An example of reachability-based acyclicity analysis}
\label{sec:runningExample}

\input running-ex-code

This section describes the essentials of the reachability-based
acyclicity analysis \cite{DBLP:conf/popl/GhiyaH96}, and its advantages
over the sharing-based one, by mean of an example.  This example will
also be used in the rest of the paper to illustrate the different
technical parts of the analysis.

Consider the program depicted in Figure~\ref{fig:run-ex-code}.  The
class \lstinline!OrderedList! implements an {\it ordered linked list}
with two fields: \lstinline!head! and \lstinline!lastInserted! point
to, respectively, the first element of the list and the last element
which has been inserted.  The class \lstinline!Node!  implements a
linked list in the standard way, with two fields \lstinline!value!
and \lstinline!next!.  Figure~\ref{fig:run-ex-diagram} shows a
possible instance of \lstinline!OrderedList!.
The method \lstinline!insert! adds a new element to the ordered list:
it takes an integer \ii, creates a new node \nn for \ii~(lines 9-10),
looks for the position \codepos of \nn~(lines 11-16), adds \nn to the
list (lines 17-20), makes \lstinline!lastInserted! point to the new
node~(lines 22), and finally returns \codepos~(lines 23).
The goal is to infer that a call of the form
``\lstinline!x.insert(i)!'' never makes \xx cyclic. This is important
since, when such call is involved in a loop like following one

\bigskip
\begin{minipage}{\textwidth}
\begin{lstlisting}
   x:=new OrderedList;
   while (j>0) do { i:=read(); x.insert(i); j:=j-1; }
\end{lstlisting}
\end{minipage}

\noindent
if \xx cannot be proven to be acyclic after \lstinline!insert!, then
it must be assumed to be cyclic from the second iteration on.  This,
in turn, prevents from proving termination of the loop at lines
$12$--$16$, since it might be traversing a cycle.

The challenge in this example is to prove that the instructions at
lines $19$ and $20$ do not make any data structure cyclic.  This is not
trivial since \codethis, \pp, and \nn share between each other at line
$17$; depending on how they share, the corresponding data structures
might become cyclic or remain acyclic.  Consider line $20$: if there
is a path (of length $0$ or more) from \nn to \pp, then the data
structures bound to them become cyclic, while they remain acyclic in
any other case.
The present analysis is able to infer that \nn and \pp share before
line $20$, but \nn does not reach \pp, which, in turn, guarantees that
no data structure ever becomes cyclic.
It can be noted that reachability information is essential for proving
acyclicity, since the mere information that \pp and \nn share, without
knowing how they do, requires to consider them as possibly cyclic, as
done, for example, by Rossignoli and Spoto~\cite{RossignoliS06}.

%% file: running-ex-code.tex

\begin{figure}[t]

\begin{minipage}{\textwidth}

\begin{lstlisting}
  class OrderedList {
    Node head, lastInserted;

    int insert(int i) {
      Node c,p,n;
      int pos; 
                                      $//~I_{7} = \emptyset$ 
      pos:=0;                         $//~I_{8} = \emptyset$ 
      n:=new Node;                    $//~I_{9} = \emptyset$ 
      n.value:=i;                     $//~I_{10} = \emptyset$ 
      c:=this.head;                   $//~I_{11} = \{\REACHES{\this}{c}\}$
      while (c!=null && c.value<i) do {
          pos:=pos+1;                 $//~I_{13} = \{\REACHES{\this}{c},\REACHES{\this}{p},\REACHES{p}{c}\}$
          p:=c;                       $//~I_{14} = \{\REACHES{\this}{c},\REACHES{\this}{p}\}$
          c:=c.next;                  $//~I_{15} = \{\REACHES{\this}{c},\REACHES{\this}{p},\REACHES{p}{c}\}$
      }                               $//~I_{16} = \{\REACHES{\this}{c},\REACHES{\this}{p},\REACHES{p}{c}\}$
      n.next:=c;                      $//~I_{17} = \{\REACHES{\this}{c},\REACHES{\this}{p},\REACHES{p}{c},\REACHES{n}{c}\}$ 
      if (p=null) 
        then this.head:=n;            $//~I_{19} = \{\REACHES{\this}{c},\REACHES{\this}{p},\REACHES{\this}{n},\REACHES{p}{c},\REACHES{n}{c}\}$ 
        else p.next:=n;               $//~I_{20} = I_{17} \cup \{\REACHES{\this}{n},\REACHES{p}{n}\}$ 
                                      $//~I_{21} = I_{19} \cup  I_{20} = I_{20}$ 
      this.lastInserted:=n;           $//~I_{22} = I_{21}$ 
      return pos+1;                   $//~I_{23} = I_{22}$ 
    }
  }

  class Node {
    Node next;
    int value;
  }
\end{lstlisting}

\end{minipage}

\caption{The running example and the result of the analysis, put in
  comments.}
  \label{fig:run-ex-code}
\end{figure}

\begin{figure}
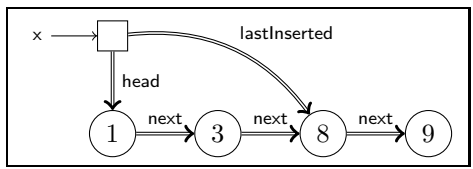

  
  \begin{center}
    \fbox{
      \input running-ex-diagram
    }
  \end{center}
  
  \caption{A graphical representation of the data structure on which the
    example works.}
  \label{fig:run-ex-diagram}
\end{figure}

%% file: running-ex-diagram.tex

\begin{tikzpicture}
  \tikzstyle{node}=[draw,circle,minimum size=4mm]
  \tikzstyle{ordl}=[draw,minimum size=4mm]
  \tikzstyle{myvar}=[minimum size=2mm]
  \tikzstyle{myderef}=[->,double]
  \node[ordl]  (xloc) at (0  ,0   ) {};
  \node[node]  (el1)  at (0  ,-1.3) {1};
  \node[node]  (el2)  at (1.4,-1.3) {3};
  \node[node]  (el3)  at (2.8,-1.3) {8};
  \node[node]  (el4)  at (4.2,-1.3) {9};
  \node[myvar] (x)    at (-1,0) {\lstinline[basicstyle=\scriptsize\sffamily]!x!};

  \draw[->] (x) -- (xloc);
  \draw[myderef] (xloc) -- node[auto] {\lstinline[basicstyle=\scriptsize\sffamily]!head!} (el1);
  \draw[myderef] (el1) -- node[auto] {\lstinline[basicstyle=\scriptsize\sffamily]!next!} (el2);
  \draw[myderef] (el2) -- node[auto] {\lstinline[basicstyle=\scriptsize\sffamily]!next!} (el3);
  \draw[myderef] (el3) -- node[auto] {\lstinline[basicstyle=\scriptsize\sffamily]!next!} (el4);
  \draw[myderef,bend left] (xloc) to node[auto] {\lstinline[basicstyle=\scriptsize\sffamily]!lastInserted!} (el3);
\end{tikzpicture}

%% file: oo-language-short.tex

\section{A simple object-oriented language}
\label{sec:language}

This section defines the syntax and the denotational semantics of a
simplified version of Java.  Class, method, field, and variable names
are taken from a set $\identifiers$ of valid \emph{identifiers}.
A \emph{program} consists of a set of classes
$\classes\subseteq\identifiers$ ordered by the \emph{subclass}
relation $\prec$.
Following Java, a \emph{class declaration} takes the form
``\lstinline!class $\class_1$ [extends $\class_2$] { $t_1~f_1$;$\ldots$ $t_n~f_n$; $M_1$ $\ldots$ $M_k$}!''
where 
each ``$t_i~f_i$'' declares the field $f_i$ to have type
$t_i\in\classes\cup\{\integer\}$,
and each $M_i$ is a method definition.
Similarly to Java, the optional statement
``\lstinline!extends $\class_2$!'' declares $\class_1$ to be a
subclass of $\class_2$.
A \emph{method definition} takes the form
``\lstinline!$t$ $m$($t_1~w_1$,$\ldots$,$t_n~w_n$) {$t_{n+1}~w_{n+1}$;$\ldots$$t_{n+p}~w_{n+p}$; $\mathit{com}$}!''
where:
$t\in\classes\cup\{\integer\}$ is the type of the return value;
$w_1,\ldots,w_n \in\identifiers$ are the formal parameters;
$w_{n+1},\ldots,w_{n+p}\in\identifiers$ are local variables; and
$\mathit{com}$ is a sequence of instructions according to the
following grammar:
\[
\begin{array}{rl}
  \mathit{exp}~{:}{:}{=} & n \mid \nil \mid  v \mid v.f \mid \mathit{exp}_1 \oplus \mathit{exp}_2 \mid \newk{\class} \mid v.m\mbox{\lstinline!(!}\bar{v}\mbox{\lstinline!)!} \\
  \mathit{com}~{:}{:}{=} & 
  v\assign \mathit{exp} \mid 
  v.f\assign \mathit{exp} \mid 
  \mathit{com_1}\mbox{\lstinline!;!}\mathit{com_2} \mid\\
  & \ifte~\mathit{exp}~\iftethen~\mathit{com_1}~\ifteelse~\mathit{com_2} \mid 
  \while~\mathit{exp}~\whilebody~\mathit{com} \mid
  \return~\mathit{exp}
\end{array} 
\]
where $v,\bar{v},m,f\in\identifiers$; $n\in\mathbb{Z}$;
$\class\in\classes$; and $\oplus$ is a binary operator (Boolean
operators return $1$ for \lstinline!true! and $0$ for
\lstinline!false!).  For simplicity, and without loss of generality,
\emph{conditions} in \ifte and \while statements are assumed not to
create objects or call methods.
A \emph{method signature} $\class.m(t_1,\ldots,t_{n}){:}t$ refers to a
method $m$ defined in class $\class$, taking $n$ parameters of type
$t_1,\ldots,t_{n} \in \classes\cup\{\integer\}$, and returning a value
of type $t$.
Given a method signature $\methodsig{m}$, let $\body{\methodsig{m}}$
be its code $\mathit{com}$; $\inp{\methodsig{m}}$ its set of input
variables $\{\this,w_1,\ldots,w_n\}$; $\locals{\methodsig{m}}$ its set
of local variables $\{w_{n+1},\ldots,w_{n+m}\}$; and
$\scope{\methodsig{m}}=\inp{\methodsig{m}}\cup\locals{\methodsig{m}}$.

A \emph{type environment} $\typenv$ is a partial map from $\variables$
to $\classes\cup\{\integer\}$ which associates types to variables at a
given program point.  Abusing notation, when the context is clear,
type environments will be confused with sets of variables; i.e., the
partial map will be confused with its domain when the type of
variables can be ignored.
A \emph{state} over $\typenv$ is a pair consisting of a frame and a
heap.
A \emph{heap} $\heap$ is a partial mapping from an infinite and
totally ordered set $\locations$ of memory locations to objects;
$\heap(\ell)$ is the object bound to $\ell\in\locations$ in the heap
$\heap$.
An \emph{object} $o \in \objects$ is a pair consisting of a class tag
$\objtag{o}\in\classes$, and a frame $\objframe{o}$ which maps its
fields into $\values=\mathbb{Z}\cup\locations\cup\{\nil\}$.
Shorthand is used: $o.f$ for $\objframe{o}(f)$; $\heap[\ell\mapsto
o]$ to modify the heap $\heap$ such that a new location $\ell$ points to
object $o$; and $\heap[\ell.f \mapsto {\rm v}]$ to modify the value of
the field $f$ of the object $\heap(\ell)$ to ${\rm v}\in\values$.
A \emph{frame} $\frm$ maps variables in $\dom(\typenv)$ to $\values$.
For $v \in \dom(\typenv)$, $\frm(v)$ refers to the value of $v$, and
$\frm[v\mapsto {\rm v}]$ is the frame where the value of $v$ has been
set to ${\rm v}$, or defined to be ${\rm v}$ if $v\not\in\dom(\frm)$.
The set of possible states over $\typenv$ is
\[
\states{\typenv}=\left\{\state{\frm}{\heap} \left|
    \begin{array}{rl}
      1. & \frm\text{ is a frame over } \typenv \text{, }
      \heap\text{ is a heap, and both are well-typed}  \\
      2. & \codom(\frm)\cap\locations\subseteq\dom(\heap)\\
      3. & \forall\ell\in\dom(\heap).~\codom(\objframe{\heap(\ell)})\cap
      \locations\subseteq\dom(\heap) \\
    \end{array}
  \right.\!\!\right\}
\]
Given $\statesym \in \states{\typenv}$, $\statef{\statesym}$ and
$\statem{\statesym}$ refer to its frame and its heap,
respectively. 
The complete lattice
$\condom=\tuple{\wp(\states{\typenv}),\states{\typenv},\emptyset,\cap,\cup}$
defines the \emph{concrete computation domain}.

A \emph{denotation} $\den$ over two type environments $\typenv_1$ and
$\typenv_2$ is a partial map from $\states{\typenv_1}$ to
$\states{\typenv_2}$: it basically describes how the state changes
when a piece of code is executed.
The set of denotations from $\typenv_1$ to $\typenv_2$ is
$\denset{\typenv_1}{\typenv_2}$.
\emph{Interpretations} are special denotations which give a meaning to
methods in terms of their input and output variables.
An interpretation $\interp\in \interpretations$ maps methods to
denotations, and is such that $\interp(\methodsig{m}) \in
\denset{\inp{\methodsig{m}}}{\{\out\}}$ for each signature
$\methodsig{m}$ in the program.
Note that the variable $\out$ is a special variable which will be used
to denote the return value of a method.
%
%

\input den-semantics.tex

Denotations for expressions and commands are depicted in
Figure~\ref{fig:den-semantics}.
An expression denotation $\einter{\typenv}{\iota}{\mathit{exp}}$ maps
states from $\states{\typenv}$ to states from
$\states{\typenv\cup\{\res\}}$, where $\res$ is a special variable for
storing the expression value.
A command denotation $\cinter{\typenv}{\iota}{com}$ maps states to
states, in presence of $\iota\in\interpretations$.
The function $\newobj{\class}$ creates a new instance of the class
$\class$ with integer fields initialized to $0$ and reference fields
initialized to $\nil$, while $\newloc{\statem{\statesym}}$ returns the
first free location, i.e., the first $\ell \notin
\dom(\statem{\statesym})$ according to the total ordering on
locations.  The function $\lookup$ resolves the method call and
returns the signature of the method to be called.
The \emph{concrete denotational semantics} of a program is defined as
the \emph{least fixpoint} of the following transformer of
interpretations~\cite{BossiGabbrielliLeviMartelli94}.

\begin{definition}
  \label{def:conc-den-semantics}
  
  The denotational semantics of a program $P$ is the least fixpoint
  (\emph{lfp}) of the following operator:
  \[
  \tp{P}(\iota) = \{ \methodsig{m}\mapsto\lambda\sigma \in \states{\inp{\methodsig{m}}}.
  \project
      {
        \cinter{\scope{\methodsig{m}}\cup\{\out\}}{\iota}{\body{\methodsig{m}}}
        (\extend{\sigma}{\methodsig{m}})
      }
      {\typenv{\setminus}\out}
      ~|~ \methodsig{m} \in P \}
      \]
      where
      $\extend{\sigma}{\methodsig{m}}=\state{\statef{\statesym}[\forall
          v\in\locals{\methodsig{m}}\cup\{\out\}. v\mapsto
          0/\nil]}{\statem{\statesym}}$.
      %
\end{definition}

\noindent
The denotation for a method signature $\methodsig{m}\in P$ is computed
by the above operator as follows:
(1) it extends (using $\extend{\sigma}{\methodsig{m}}$) the input
state $\statesym \in \states{\inp{\methodsig{m}}}$ such that local
variables are set to $0$ or \nil, depending on their type;
(2) it computes the denotation of the code of \methodsig{m} (using
$\cinter{\scope{\methodsig{m}}\cup\{\out\}}{\iota}{\body{\methodsig{m}}}$); and
(3) it restricts the resulting denotation to the output variable
$\out$ (using $\exists\typenv{\setminus}\out$).

%% file: den-semantics.tex

\begin{figure}[t]

\begin{center}
\begin{minipage}{\textwidth}
\[
\small
\begin{array}{@{}r@{}l@{}}
  \einter{\typenv}{\interp}{n}(\statesym) =& 
     \state{\statef{\statesym}[\res\mapsto n]}{\statem{\statesym}} \\
  \einter{\typenv}{\interp}{\nil}(\statesym) =& 
     \state{\statef{\statesym}[\res\mapsto \nil]}{\statem{\statesym}} \\
  \einter{\typenv}{\interp}{\newk{\class}}(\statesym) =& 
     \state{\statef{\statesym}[\res \seq \ell]}
           {\statem{\statesym}[\ell \seq \newobj{\class}]}
      \mbox{ where } \ell = \newloc{\statem{\statesym}} \\
  \einter{\typenv}{\interp}{v}(\statesym) =& 
     \state{\statef{\statesym}[\res\mapsto \statef{\statesym}(v)]}{\statem{\statesym}}\\
  \einter{\typenv}{\interp}{v.f}(\statesym) =& 
     \state{\statef{\statesym}[\res\mapsto \statem{\statesym}(\statef{\statesym}(v)).f]}{\statem{\statesym}}\\
  \einter{\typenv}{\interp}{\mathit{exp}_1{\oplus}\mathit{exp}_2}(\statesym) =&  
     \state{\statef{\statesym}[\res\mapsto \statef{\statesym}_1(\res)  \oplus  \statef{\statesym}_2(\res)]}
           {\statem{\statesym}_2}\mbox{ where } \\
  & ~~~~\statesym_1=\einter{\typenv}{\interp}{\mathit{exp}_1}(\statesym) \mbox{ and } \statesym_2=\einter{\typenv}{\interp}{\mathit{exp}_1}(\state{\statef{\statesym}}{\statem{\statesym}_1})\\
  \einter{\typenv}{\interp}{v_0.m(v_1,\ldots,v_n)}(\statesym) =&
   \state{\statef{\statesym}[\res \seq \statef{\statesym}_2(out)]}{\statem{\statesym}_2}
   \mbox{ where }  \\
       &~~~~ \statesym_2=\interp(\methodsig{m})(\statesym_1) \mbox { and } \statesym_1 \mbox{ is such that }\\ 
       &~~~~~~ 1.~ \statem{\statesym}_1 = \statem{\statesym}; \\
       &~~~~~~ 2.~\statef{\statesym}_1(this)=\statef{\statesym}(v_0);\\
       &~~~~~~ 3.~\forall 1{\le}i{\le}n.~\statef{\statesym}_1(w_i)=\statef{\statesym}(v_i); \mbox{ and}\\
       &~~~~~~ 4.~\methodsig{m} = \lookup(\statesym,v_0.m(v_1,\ldots,v_n)); 
  \\
\hline
  %
  %
  \cinter{\typenv}{\interp}{v\assign\mathit{exp}}(\statesym) =&
     \state{\statef{\statesym}[v \mapsto \statef{\statesym}_e(\res)]}
           {\statem{\statesym}_e} 
  \\
  %
  %
  \cinter{\typenv}{\interp}{v.f\assign\mathit{exp}}(\statesym) =&
     \state{\statef{\statesym}}
           {\statem{\statesym}[\ell.f \mapsto \statef{\statesym}_e(\res)]}
           \mbox{ where } \ell=\statef{\statesym}(v)
  \\
  %
  %
  \cinter{\typenv}{\interp}{
    \begin{array}{@{}r@{}l@{}}
      \ifte~\mathit{exp}~ & \iftethen~\mathit{com}_1 \\
                          & \ifteelse~\mathit{com}_2 
    \end{array}}(\statesym)
  =& 
   \mbox{if $\statef{\statesym}_e(\res) \neq 0$
               then $\cinter{\typenv}{\interp}{\mathit{com}_1}(\statesym)$ 
               else $\cinter{\typenv}{\interp}{\mathit{com}_2}(\statesym)$ }
  \\
  %
  %
  \cinter{\typenv}{\interp}{\while~\mathit{exp}~\whilebody~\mathit{com}}(\statesym)
  =& \den(\statesym) \mbox{ where } \den \mbox{ is the least fixpoint
    of }\\
  &   \mbox{$\lambda w.\lambda\statesym.$ 
    if $\statef{\statesym}_e(\res) \neq 0$ 
    then $w(\cinter{\typenv}{\interp}{\mathit{com}}(\statesym))$ else $\statesym$}
  \\
  %
  %
  \cinter{\typenv}{\interp}{\return~\mathit{exp}}(\statesym) =& 
    \state{\statef{\statesym}[\out \mapsto \statef{\statesym}_e(\res)]}
           {\statem{\statesym}_e} 
  \\
  %
  %
  \cinter{\typenv}{\interp}{\mathit{com}_1;\mathit{com}_2}(\statesym) =&
     \cinter{\typenv}{\interp}{\mathit{com}_2}(\cinter{\typenv}{\interp}{\mathit{com}_1}(\statesym))
  \\
\end{array}
\]
\end{minipage}
\end{center}
\caption{Denotations for expressions and commands.  The state
  $\statesym_e$ is $\einter{\typenv}{\interp}{\mathit{exp}}(\statesym)$.}
\label{fig:den-semantics}
\end{figure}

%% file: domain.tex

\section{The abstract domain}
\label{sec:abstractDomain}

The acyclicity analysis discussed in this paper works on the reduced
product~\cite{Cousot79} of two \emph{abstract domains}, according to
the theory of Abstract Interpretation~\cite{Cousot77}.
The first domain captures \emph{may-reachability}, while the second
deals with the \emph{may-be-cyclic} property of variables.
Both are based on the notion of \emph{reachable heap locations}, i.e.,
the part of the heap which can be reached from a location by accessing
object fields.

\begin{definition}[reachable heap locations~\cite{RossignoliS06}]
  \label{def:reachability}
  Given a heap $\heap$, the set of reachable locations from
  $\ell\in\dom(\heap)$ is $\reachable{\heap}{\ell}=\cup\{
  \reachablei{i}{\heap}{\ell} \mid i \ge 0 \}$,
  where
  $\reachablei{0}{\heap}{\ell}=
  \codom(\objframe{\heap(\ell)})\cap\locations$, and
  $\reachablei{i+1}{\heap}{\ell}=\cup\{\codom(\objframe{\heap(\ell')})
  \cap \locations \mid \ell' \in \reachablei{i}{\heap}{\ell}\}$.
  The set of $\varepsilon$-reachable locations from
  $\ell\in\dom(\heap)$ is
  $\ereachable{\heap}{\ell}=\reachable{\heap}{\ell}\cup\{\ell\}$.
\end{definition}

\noindent
Note that $\varepsilon$-reachable locations include the source
location $\ell$ itself, while reachable locations do not (unless
$\ell$ is reachable from itself through a \emph{cycle} whose length is
at least 1).
The rest of this section is developed in the context of a given type
environment $\typenv$.

\subsection{Reachability}
Given a state $\statesym\in\states{\typenv}$, a reference variable
$v\in\typenv$ is said to \emph{reach} a reference variable $w\in\typenv$ in $\statesym$ if
$\statef{\statesym}(w) \in
\reachable{\statem{\statesym}}{\statef{\statesym}(v)}$.
This means that, starting from $v$ and applying \emph{at least one
  dereference operation} (i.e., going from the location pointed to by $v$
to the location pointed to by $v.f$ for some field
$f$), it is possible to reach the object to which $w$
points.
Due to strong typing, $\typenv$ puts some restrictions on
reachability; i.e., it might be impossible to have a heap where a
variable of type $\class_1$ reaches one of type $\class_2$.
Following Secci and Spoto \cite{spoto:pair_sharing}, a class
$\class_2\in\classes$ is said to be \emph{reachable} from
$\class_1\in\classes$ if there exist $\statesym\in\states{\typenv}$,
and two locations $\ell,\ell'\in\dom(\statem{\statesym})$ such
that~(a) $\objtag{\statem{\statesym}(\ell)} = \class_1$; (b)
$\objtag{\statem{\statesym}(\ell')} = \class_2$; and (c) $\ell' \in
\reachable{\statem{\statesym}}{\ell}$.  The use of this notion (as
well as the notion of cyclic class introduced in Section
\ref{sec:cyclicity} and used in Definition \ref{def:abs-dom-cyc}) in
the definition of the reachability and cyclicity domains allows us to
obtain the needed Galois insertions.  It must be pointed out that both
notions can be computed statically, so that they can be assumed to be
pre-computed information.


\begin{definition}[reachability domain]
  \label{def:abs-dom-reach}
  The reachability abstract domain is the complete lattice
  $\domr=\tuple{\wp(\reachset{\typenv}),\subseteq,
    \emptyset,\cR^\typenv,\cap,\cup}$, where
  \[
  \cR^\typenv = \left\{\REACHES{v}{w} ~\left| \begin{array}{l} v, w
    \in \dom(\typenv)\mbox{, and there exist
      $\class_1{\subclasseq}\typenv(v)$ and
      $\class_2{\subclasseq}\typenv(w)$} \\ \mbox{such that
      $\class_2$ is reachable from $\class_1$} \end{array}
  \right. \right\}
  \]
  Here and in the following, elements of the tuple
  $\tuple{A,\leq,\bot,\top,\wedge,\vee}$ denoting an abstract domain
  $\mathcal{A}$ represent, respectively, ($A$) the set of abstract
  values, ($\leq$) the partial order on them, ($\bot$) the minimal
  (bottom) element of $A$, ($\top$) the maximal (top) element of $A$,
  ($\wedge$) the meet operator and ($\vee$) the join operator on $A$.
  This terminology is standard in Abstract Interpretation.
\end{definition}

\noindent 
\emph{May-reach} information is described by \emph{abstract values}
$\abselemr \in \wp(\reachset{\typenv})$.  For example,
$\{\REACHES{\xx}{\zz},\REACHES{\yy}{\zz}\}$ describes those states
where \xx and \yy \emph{may} reach \zz.  Note that a statement
$\REACHES{\xx}{\yy}$ does not prevent \xx and \yy from aliasing;
instead, \xx can reach \yy and alias with it at the same time, e.g.,
when \xx, \yy, and \lstinline!x.f! point to the same location.

\begin{lemma}
  \label{lemma:gi-reach}
  The following abstraction and concretization functions define a
  \emph{Galois insertion} between $\domr$ and $\condom$:
  \[
  \begin{array}{ll}
    \alphar(\concelem) = & \{ \REACHES{v}{w}\in\reachset{\typenv}
    ~|~ \exists \statesym\in
    \concelem. v~\mbox{reaches}~w~\mbox{in}~\statesym\} \\
    \gammar(\abselemr) = & \{ \statesym\in\states{\typenv} ~|~ ~\forall
    v,w\in\typenv.~ v~\mbox{reaches}~w~\mbox{in}~\statesym
    \Rightarrow \REACHES{v}{w}\in \abselemr \}  \\
  \end{array}
  \]
\end{lemma}

\noindent
The top element $\reachset{\typenv}$ is $\alphar(\states{\typenv})$,
and represents all states which are compatible with $\typenv$.  This
is because the presence of a reachability statement in an abstract
value $I$ does not require a reachability path to actually exist;
rather, the concretization of $I$ will include states where the path
does exist, and states where it does not (this is the meaning of
``may-information''). 
In other words, the absence of a reachability statement in the
abstract state requires non-existence of a reachability path in its
concretization.

The bottom element $\emptyset$ models the set of all states where, for
every two reference variables $v$ and $w$ (possibly the same
variable), $v$ does not reach $w$.  Note that, clearly, this set is
not empty, and that the absence of a reachability statement actually
rules out states where the reachability path exists.

\begin{remark}
  \label{remark:transitivity}
  Intuitively, reachability is a transitive property; i.e., if \xx
  reaches \yy and \yy reaches \zz, then \xx also reaches \zz.
  However, values in $\domr$ are \emph{not} closed by transitivity:
  e.g., it is possible to have $I_r = \{ \REACHES{\xx}{\yy},
  \REACHES{\yy}{\zz} \}$ which contains $\REACHES{\xx}{\yy}$ and
  $\REACHES{\yy}{\zz}$, but not $\REACHES{\xx}{\zz}$.  Such an
  abstract value is a reasonable one, and approximates, for example,
  the execution of the following code

  \bigskip
  \begin{minipage}{\textwidth}
    \begin{lstlisting}
      x:=new C; 
      y:=new C; 
      if (w$>$0) then x.f:=y; else y.f:=z;
    \end{lstlisting}
  \end{minipage}
  
  \noindent
  Moreover, this abstract value is consistent, i.e., it describes a
  set of concrete states which is not smaller (actually, it is
  greater) than $\gammar(\emptyset)$.  This happens because
  reachability is, actually, \emph{may-reach} information, so that,
  for example, $\gammar(\{ \REACHES{\xx}{\yy}, \REACHES{\yy}{\zz} \})$
  includes (a) any state where \xx reaches \yy but \yy does not reach
  \zz; (b) any state where \yy reaches \zz but \xx does not reach \yy;
  and (c) any state where \xx does not reach \yy and \yy does not
  reach \zz.
  It is important to point out that $\gammar(\{ \REACHES{\xx}{\yy},
  \REACHES{\yy}{\zz} \})$ does not contain those states where both \xx
  reaches \yy and \yy reaches \zz, since, in this case, \xx would also
  reach \zz by transitivity, which is forbidden by soundness since
  $\REACHES{\xx}{\zz} \notin \abselemr$.
\end{remark}

\subsection{Cyclicity}
\label{sec:cyclicity}

Given a state $\statesym\in\states{\typenv}$, a variable
$v\in\dom(\typenv)$ is said to be \emph{cyclic} in $\statesym$ if
there exists
$\ell\in\ereachable{\statem{\statesym}}{\statef{\statesym}(v)}$ such
that $\ell\in\reachable{\statem{\statesym}}{\ell}$.  In other words,
$v$ is cyclic if it reaches some memory location $\ell$ (which can
possibly be $\statef{\statesym}(v)$ itself) through which a cyclic
path goes.
Similarly to reachability, it might be impossible to generate a cyclic
data structure starting from a variable of some type $\class$.  A
class $\class\in\classes$ is said to be a \emph{cyclic class} if there
exist $\statesym\in\states{\typenv}$ and
$\ell,\ell'\in\dom(\statem{\statesym})$ such that
$\objtag{\statem{\statesym}(\ell)}=\class$,
$\ell'\in\ereachable{\statem{\statesym}}{\ell}$, and
$\ell'\in\reachable{\statem{\statesym}}{\ell'}$.
The cyclicity domain is the dual of the non-cyclicity domain by
Rossignoli and Spoto \cite{RossignoliS06}.


\begin{definition}[cyclicity domain]
  \label{def:abs-dom-cyc}
  The abstract domain for cyclicity is represented as the complete
  lattice $\domc=\tuple{\wp(\cycset{\typenv}),\subseteq,\emptyset,
    \cycset{\typenv},\cap,\cup}$ where
  \[
  \cycset{\typenv} = \{ \CYCLE{v}~|~v \in \typenv, \mbox{and there
    exists a cyclic class}~\class \subclasseq \typenv(v) \}
  \]
\end{definition}

\begin{lemma}
  \label{lemma:gi-cyc}
  The following abstraction and concretization functions define a
  Galois insertion between $\domc$ and $\condom$
  \[
  \begin{array}{ll} 
    \alphac(\concelem)   = & \{ \CYCLE{v} ~|~ \exists v\in\typenv.~ \exists
    \statesym\in \concelem.~ v \mbox{ is cyclic in } \statesym\}\\
    \gammac(\abselemc) = & \{ \statesym ~|~ \statesym\in\states{\typenv} \wedge
    \forall v\in \typenv.~ (v~\mbox{is cyclic in}~\statesym) \Rightarrow
    \CYCLE{v} \in \abselemc\}
  \end{array}  
  \]
\end{lemma}

\noindent 
\emph{May-be-cyclic} information is described by \emph{abstract
  values} $\abselemc \in \wp(\cycset{\typenv})$.  For instance, $\{
\CYCLE{\xx} \}$ represents states where no variable but \xx can be
cyclic.  The top element $\cycset{\typenv}$ is concretized to
$\states{\tau}$; i.e., all state are included since each variable can
be either cyclic or acyclic.  The bottom element $\emptyset$ does not
allow any variable to be cyclic, i.e., its concretization does not
include any state with cyclic variables.

\subsection{The reduced product}

As it will be explained in Section~\ref{sec:abstractSemantics}, the
abstract semantics uses reachability information in order to detect
cycles, and cyclicity information in order to add, in some cases,
reachability statements.  Both kinds of information can be combined:
in the theory of Abstract Interpretation, this amounts to computing
the \emph{reduced product} \cite{Cousot79} of the corresponding
abstract domains.  In the present context, the reduced product is
obtained by \emph{reducing} the Cartesian product
$\domrc{\typenv}=\domr\times\domc$.  Elements of $\domrc{\typenv}$ are
pairs $\tuple{\abselemr,\abselemc}$, where $\abselemr$ and $\abselemc$
contain, respectively, the may-reach and the may-be-cyclic
information.
The abstraction and concretization functions are induced by those on
$\domc$ and $\domr$:
\[
\begin{array}{ccc}
  \gammarc(\tuple{\abselemr,\abselemc}) = \gammar(\abselemr) \cap \gammac(\abselemc) 
  &~~~~&
  \alpharc(I) = \tuple{\alphar(I),\alphac(I)}\\
\end{array}
\]
However, it can happen that two elements of $\domrc{\typenv}$ are
mapped to the same set of concrete elements, which prevents having a
Galois insertion between $\domrc{\typenv}$ and $\condom$.  The
operation of reduction deals exactly with this problem.
In order to compute it, an equivalence relation $\equiv$ has to be
defined, which satisfies $I_{rc}^1\equiv I_{rc}^2$ is and only if
$\gammarc(I_{rc}^1)=\gammarc(I_{rc}^2)$.  Functions $\gammarc$ and
$\alpharc$ define a Galois insertion between
${\domrc{\typenv}}_\equiv$ and $\condom$, where
${\domrc{\typenv}}_\equiv$ is $\domrc{\typenv}$ equipped (reduced)
with the equivalence relation.
The following lemma characterizes the equivalence relation on
$\domrc{\typenv}$.

\begin{lemma}
  \label{lemma:equivalenceClass}
  For any abstract values $\abselemr^1,\abselemr^2\in\domr$ and
  $\abselemc^1,\abselemc^2\in\domc$, the concretization
  $\gammarc(\tuple{\abselemr^1,\abselemc^1})$ is equal to
  $\gammarc(\tuple{\abselemr^2,\abselemc^2})$ if and only if both
  conditions hold: (a) $\abselemc^1 = \abselemc^2$; and (b)
  $\abselemr^1 \setminus \{ \REACHES{v}{v}~|~\CYCLE{v}\notin
  \abselemc^1 \} = \abselemr^2 \setminus \{
  \REACHES{v}{v}~|~\CYCLE{v}\notin \abselemc^2 \}$.
\end{lemma}

\noindent
This above lemma means that: (a) may-be-cyclic information always
makes a difference as regards the set of concrete states; that is,
adding a new statement $\CYCLE{v}$ to $\abselemrc\in\domrc{\typenv}$
results in representing a strictly larger set of states; and (b)
adding a pair $\REACHES{v}{v}$ to $\abselemrc\in\domrc{\typenv}$, when
$v$ cannot be cyclic, does not make it represent more concrete states,
since the acyclicity of $v$ excludes that it can reach itself.

\begin{example}
  \label{ex:red-1}
  As an example for case (a), consider two abstract values
  $\abselemrc^1 = \tuple{\abselemr,\emptyset}$ and $\abselemrc^2 =
  \tuple{\abselemr, \{\CYCLE{\xx}\}}$ which result from adding
  $\CYCLE{\xx}$ to $\abselemrc^1$.
  Assuming that \xx does not appear in $\abselemr$, there is a state
  $\statesym$ which is compatible with $\abselemr$ (for example, if no
  $v$ reaches any $w$ in $\statesym$), and where \xx is cyclic (note
  that this does not require \xx to reach any other variable, not even
  itself, since the cycle does not need to go through
  $\statef{\statesym}(\xx)$).
  This $\statesym$ belongs to $\gammarc(\abselemrc^2) \setminus
  \gammarc(\abselemrc^1)$ and is, therefore, an example of the
  difference between the abstract values.
  
  As an example for (b), consider $\abselemrc^1 =
  \tuple{\emptyset,\{\CYCLE{\yy}\}}$ and $\abselemrc^2 =
  \tuple{\{\REACHES{\xx}{\xx}\},\{\CYCLE{\yy}\}}$ which results from
  adding $\REACHES{\xx}{\xx}$ to $\abselemrc^1$.  At a first glance,
  $\abselemrc^2$ describes a larger set of states, since it includes
  states (not belonging to $\gammarc(\abselemrc^1)$) where there is a
  path from \xx to \xx.  However, such states will neither belong to
  $\gammarc(\abselemrc^2)$, since such a path implies that \xx is
  cyclic, which is not permitted by $\{\CYCLE{\yy}\}$, that only
  allows \yy to be cyclic.
\end{example}

\noindent
Lemma~\ref{lemma:equivalenceClass} provides a way for computing the
\emph{normal form} of any $\tuple{\abselemr,\abselemc}$, which comes
to be $\tuple{\abselemr \setminus \{ \REACHES{v}{v}|\CYCLE{v}{\notin}
  \abselemc \},\abselemc}$, i.e., the \emph{canonical form} of its
equivalence class.
From now on, $\domrc{\typenv}$ will be a shorthand for
${\domrc{\typenv}}_\equiv$, where $\equiv$ is left implicit.

%% file: analysis-short.tex

\section{Reachability-based acyclicity analysis}
\label{sec:abstractSemantics}

This section uses $\domrc{\typenv}$ to define an abstract semantics
from which one can decide whether a variable $v$ is (or may not be)
bounded to an acyclic data structure at a given program point.
Informally, two variables $v$ and $w$ are said to \emph{share} in a
state $\statesym$ if they $\varepsilon$-reach (i.e., in zero or more
steps) a common location in the heap.  The analysis is based on the
observation that reachability information can tell \emph{how} $v$ and
$w$ share: this can happen because either (a) $v$ and $w$ alias; (b)
$v$ reaches $w$; (c) $w$ reaches $v$; or (d) they both reach $\ell \in
\dom(\statem{\statesym})$.
Distinguishing among these four possibilities is crucial for a precise
acyclicity analysis.  In fact, assuming that $v$ and $w$ are initially
acyclic, they both become cyclic after executing
\lstinline!$v$.f:=$w$!  \emph{if and only if}, initially, $w$ either
reaches $v$ or aliases with it.  This is clearly more precise than
declaring $v$ as cyclic whenever it was sharing with
$w$~\cite{RossignoliS06}.
The presented analysis is an adaptation of the work by Ghiya and
Hendren~\cite{DBLP:conf/popl/GhiyaH96} to an object-oriented
framework, where the chosen formalism is that of an abstract semantics
on the domain described in Section \ref{sec:abstractDomain}.  Some
optimizations w.r.t.~the original analysis are also discussed.

The rest of this section formalizes the reachability-based analysis as
an \emph{abstract semantics} on $\domrc{\typenv}$, and proves some
important results.

\subsection{Preliminaries}
\label{sec:preliminaries}

\emph{May-share} \cite{spoto:pair_sharing}, \emph{may-alias}
\cite{Hind01} and \emph{purity} \cite{GenaimS08} analyses are used as
pre-existent components, i.e., programs are assumed to have been
analyzed w.r.t.\ these properties by means of state-of-the-art
tools\footnote{One could argue that aliasing and sharing analyses
  benefit from reachability information, so that all the components
  should better work ``in parallel''; however, for the sake of this
  presentation, the three components (sharing, aliasing, and
  reachability-cyclicity) are supposed to be independent.  See Section
  \ref{sec:completeness} for further discussion about the interplay
  between all the analyses.}.  Two reference variables $v$ and $w$
\emph{share} in $\statesym$ iff
$\ereachable{\statem{\statesym}}{\statef{\statesym}(v)}\cap\ereachable{\statem{\statesym}}{\statef{\statesym}(w)}
\neq\emptyset$; also, they \emph{alias} in $\statesym$ if they point
to the same location, namely, if
$\statef{\statesym}(v)=\statef{\statesym}(w)\in\dom(\statem{\statesym})$.
Any non-null reference variable shares and aliases with itself; also,
both are symmetric relations.
The $i$-th argument of a method $\methodsig{m}$ is said to be
\emph{pure} if $\methodsig{m}$ does not update the data structure to
which the argument initially points.
For \emph{sharing} and \emph{purity}, the analysis proposed by Genaim
and Spoto~\cite{GenaimS08} (based on previous work by Secci and
Spoto~\cite{spoto:pair_sharing}) can be applied: with it,
\begin{enumerate}
\item it is possible to know if $v$ may share with $w$ at any program
  point (denoted by the sharing statement $\SHARE{v}{w}$); and
\item for each method $\methodsig{m}$, a denotation
  $\shden{\methodsig{m}}$ is given: for a set of pairs $I_{sp}$ which
  safely describes the sharing between actual arguments in the input
  state, $I'_{sp}=\shden{\methodsig{m}}(I_{sp})$ is such that (i) if
  $\SHARE{v}{w}\in I'_{sp}$, then $v$ and $w$ might share during the
  execution of $\methodsig{m}$; and (ii) $\PURE{v}_i\in I'_{sp}$ means
  that the $i$-th argument might be non-pure.
\end{enumerate}

\noindent
According to the theory of Abstract Interpretation and to previous
work, sharing and purity analysis can be defined as an abstract
semantics over the abstract domain $\cI_{sp}^{\typenv}$, whose
elements $I_{sp}$ contain may-share statements $\SHARE{v}{w}$ and
may-be-non-pure statements $\PURE{v}$.  Abstraction and concretization
functions $\alpha_{sp}$ and $\gamma_{sp}$ are defined in the standard
way \cite{GenaimS08}: in particular, $\gamma_{sp}(I_{sp})$ contains
all the states where variables mentioned in sharing statements are the
only ones which can possibly share between themselves, while variables
mentioned in may-be-non-pure statements are the only ones which can
possible be non-pure.

As for \emph{aliasing}, the abstract domain $\cI_{al}^{\typenv}$
contains sets of may-alias statements $\ALIAS{v}{w}$: if
$\ALIAS{v}{w}$ is contained in $I_{al}$, then its concretization
$\gamma_{al}(I_{al})$ contains states where $v$ and $w$ actually alias
and states where they do not.  It is assumed that this information is
available at each program point as a set of may-alias statements.

In the following, the domain $\cI_{s}^{\typenv}$ will be the reduced
product between $\cI_{sp}^{\typenv}$ and $\cI_{al}^{\typenv}$, and
combines sharing, aliasing, and purity information.  As usual,
$\gamma_s((I_{sp},I_{al}))$ is defined as $\gamma_{sp}(I_{sp}) \cap
\gamma_{al}(I_{al})$, while $\alpha_s(X)$ is defined as
$(\alpha_{sh}(X) \cup \alpha_{al}(X))_\equiv$, where $\equiv$ means
that abstract elements with the same concretization have been unified
(i.e., the product has been reduced).

Abusing notation, from now on $I_s$ will be often used to denote an
abstract value without specifying the abstract domain it belongs to.
The use of $I_s$ will be clear from the context: for example, writing
$\gamma_{al}(I_s)$ means applying $\gamma_{al}$ to the part of $I_s$
which represents aliasing information.

Moreover, an abstract element
$\tuple{\abselemr,\abselemc}\in\domrc{\typenv}$ will be represented by
the set $I=\abselemr\cup\abselemc$; therefore, $\REACHES{v}{w} \in I$
and $\CYCLE{v} \in I$ are shorthands for, respectively,
$\REACHES{v}{w} \in I_r$ and $\CYCLE{v} \in I_c$.
The operation $\WO{I}{v}$ (\emph{projection}) removes any statement
about $v$ from $I$, while $I[v/w]$ (\emph{renaming}) $v$ to $w$ in
$I$.
For the sake of simplicity, class-reachability and class-cyclicity are
taken into account \emph{implicitly}: a new statement $\REACHES{v}{w}$
is not added to an abstract state if
$\REACHES{v}{w}\not\in\reachset{\typenv}$, while a statement
$\CYCLE{v}$ is not added if $\CYCLE{v}\not\in\cycset{\typenv}$.  It is
important to point out that information about class-reachability and
class-cyclicity (i.e., whether $\class_1$ reaches $\class_2$, or
whether $\class$ is cyclic) can be computed statically and before
performing any acyclicity analysis.  Therefore, it can be assumed that
such information is available whenever it is necessary to decide
whether a new reachability or cyclicity statement belongs or not to
$\reachset{\typenv}$ or $\cycset{\typenv}$.

\input abs-semantics

\subsection{The abstract semantics}

An \emph{abstract denotation} $\absden$ from $\typenv_1$ to
$\typenv_2$ is a partial map from $\domrc{\typenv_1}$ to
$\domrc{\typenv_2}$.  It describes how the abstract input state
changes when a piece of code is executed.  The set of all abstract
denotations from $\typenv_1$ to $\typenv_2$ is denoted by
$\absdenset{\typenv_1}{\typenv_2}$.
As in the concrete setting, interpretations provide abstract
denotations for methods in terms of their input and output arguments.
An \emph{interpretation} $\absinterp$ maps methods to abstract
denotations, and is such that $\absinterp(\methodsig{m}) \in
\absdenset{\inp{\methodsig{m}}}{\inp{\methodsig{m}}\cup\{\out\}}$.
Note that the range of such denotations is
$\inp{\methodsig{m}}\cup\{\out\}$, instead of $\{\out\}$ (as in the
concrete semantics): this point will get clarified below.
Finally, $\absinterpretations$ denotes the set of all (abstract)
interpretations.

Figures~\ref{fig:abs-sem-exp} and \ref{fig:abs-sem-com} depict
abstract denotations.
An \emph{expression denotation}
$\EXPASEMANTICS{\typenv}{\absinterp}{\mathit{exp}}$ maps abstract
states from $\domrc{\typenv}$ to abstract states from
$\domrc{\typenv_1}$ where $\typenv_1=\typenv\cup\{\res\}$, while a
\emph{command denotation} $\ASEMANTICS{\typenv}{\absinterp}{com}$ maps
$\domrc{\typenv}$ to $\domrc{\typenv}$.
%

In the definition, the abstract element $I_s$ contains the sharing,
aliasing, and purity information pre-computed by other analyses, and
referring to the program point of interest\footnote{Note that $I_s$
  could be represented explicitly as an input to the abstract
  semantics, next to $I$, but it is not written for better clarity}.

\subsubsection{Expressions}

An expression denotation
$\EXPASEMANTICS{\typenv}{\absinterp}{\mathit{exp}}$ adds to an input
state $I$ those reachability and cyclicity statements which result
from evaluating $\mathit{exp}$.

Nothing is added to $I$ in cases $(1_e){:}~
\EXPASEMANTICS{\typenv}{\absinterp}{n}$, $(2_e){:}~
\EXPASEMANTICS{\typenv}{\absinterp}{\nil}$, and $(3_e){:}~
\EXPASEMANTICS{\typenv}{\absinterp}{\newk{\class}}$ since the
expression is evaluated without side effects to, respectively, an
integer value, \nil, or a newly allocated object which is not related
to any other location with respect to reachability.

The same reasoning explains why the returned abstract value is also
$I$ in case $(4_e){:}~\EXPASEMANTICS{\typenv}{\absinterp}{v}$ when
$\typenv(v){=}\integer$, and
$(5_e){:}~\EXPASEMANTICS{\typenv}{\absinterp}{v.f}$ when $f$ is an
\integer field.

In case $(4_e){:}~\EXPASEMANTICS{\typenv}{\absinterp}{v}$, when the
type of $v$ is not \integer, the result variable $\res$ has the same
abstract behavior as $v$.  Therefore, the semantics returns $I$,
together with a \emph{cloned} version $I[v/\res]$ where statements
about $v$ have been replaced by renamed statements about $\res$.

In the case of $(5_e){:}~\EXPASEMANTICS{\typenv}{\absinterp}{v.f}$,
  when $f$ is a reference field, the following information is added to
  $I$:
\begin{itemize}
\item statements for $v$ which are cloned for $\res$;
\item $\REACHES{w}{\res}$, if $w$ might share with $v$; note that
  $\REACHES{v}{\res}$ is always added since $\SHARE{v}{v} \in I_s$
  (clearly, $v$ cannot be \nil); if $v$ and $w$ reach a common
  location (which implies that they share), but do not reach each
  other, then, conservatively, the reachability statement
  $\REACHES{w}{\res}$ must be added because $v.f$ could be exactly the
  common location which is reached by both $v$ and $w$;
\item if $v$ might be cyclic, then, for soundness,
  $\REACHES{\res}{\res}$; note that, in this case, $\CYCLE{\res}$ is
  also guaranteed to have been previously added to the abstract state.
\end{itemize}

In case $(6_e){:}~\EXPASEMANTICS{\typenv}{\absinterp}{\mathit{exp}_1
  \oplus \mathit{exp}_2}$, the expression $\mathit{exp}_1$ is first
analyzed, then $\mathit{exp}_2$ is analyzed on the resulting abstract
state.  Note that, in both cases, $\res$ is removed since the return
value has always type $\integer$.

Finally, method calls
$(7_e){:}~\EXPASEMANTICS{\typenv}{\absinterp}{v_0.m(v_1,..,v_n)}$
will be explained later, after introducing denotations for commands.

\begin{example}
  \label{ex:analysis-exp}
  Consider \lstinline!c:=c.next! at line 15 in
  Figure~\ref{fig:run-ex-code}.  Evaluating the denotation
  $\EXPASEMANTICS{\typenv}{\absinterp}{\mbox{\lstinline!c.next!}}(I_{14})$
  results in $\{\REACHES{\this}{c}, \REACHES{\this}{p},
  \REACHES{\this}{\res}, \REACHES{c}{\res}, \REACHES{p}{\res}\}$. The
  statement $\REACHES{\this}{\res}$ is added since
  $\REACHES{\this}{c}\in I_{14}$; $\REACHES{c}{\res}$ and
  $\REACHES{p}{\res}$ are added because $\SHARE{c}{c}$ and
  $\SHARE{c}{p}$ hold after line 14.
\end{example}

\subsubsection{Variable assignment}

The denotation $(1_c){:}~\ASEMANTICS{\typenv}{\absinterp}{v\assign
  \mathit{exp}}$ computes
$\EXPASEMANTICS{\typenv}{\absinterp}{\mathit{exp}}(I)$, removes any
statement about $v$ since it takes a new value, and finally renames
$\res$ to $v$. Note that it is safe to remove statements about $v$
since it is first cloned to $\res$.

\begin{example}
  \label{ex:analysis-var-assign}
  Consider, again, line 15 in Figure~\ref{fig:run-ex-code}.
  Evaluating the denotation
  $\ASEMANTICS{\typenv}{\absinterp}{\mbox{\lstinline!c:=c.next!}}(I_{14})$
  first computes
  $\EXPASEMANTICS{\typenv}{\absinterp}{\mbox{\lstinline!c.next!}}(I_{14})$
  as in Example~\ref{ex:analysis-exp}.
  Then, statements involving $c$ are removed, which results in
  $\{\REACHES{\this}{p},\REACHES{\this}{\res},\REACHES{p}{\res}\}$,
  and, finally, $\res$ is renamed to $c$, giving
  $\{\REACHES{\this}{p},\REACHES{\this}{c},\REACHES{p}{c}\}$.
  Note that $\REACHES{\this}{c}$ is reinserted (by renaming
  $\REACHES{\this}{\res}$) after being deleted by $\exists c$. Also,
  note that $\REACHES{c}{\res}$ has been removed by $\exists c$, so
  that, correctly, $c$ is not considered to reach itself after the
  assignment.
\end{example}

\subsubsection{Field update}
\label{sec:fieldUpdate}

The denotation $(2_c){:}~\ASEMANTICS{\typenv}{\absinterp}{v.f\assign
  \mathit{exp}}$ accounts for field updates.  The set $I'_0$ results
from computing $\EXPASEMANTICS{\typenv}{\absinterp}{\mathit{exp}}(I)$,
as usual.  The following step is to apply an \emph{optimization}
(called the \emph{single-field optimization} in the following) which
allows removing statements after inspecting the declarations of the
classes involved in the update.  The abstract value $I'$ is computed
from $I'_0$ by the function $\mathit{condRemove}(I'_0,v,f)$, which is
defined as follows.  Let $\class$ be the declared class of $v$ (this
means that the runtime type of $v$ can be $\class$ or any of its
subclasses); then $I'$ is obtained by $I'_0$ by
\begin{itemize}
\item removing $\CYCLE{v}$ if (1) $f$ is the only reference field of
  any $\class'\subclasseq \class$; or (2) all the other reference
  fields of any $\class'\subclasseq \class$ have a declared class such
  that neither it nor any of its subclasses are a cyclic class;
\item similarly, removing any statement $\REACHES{v}{w}$ such that $f$
  is the only field of any $\class'\subclasseq \class$ whose declared
  class $\class_f$ (or any of its subclasses) reaches the declared
  class of $w$ (or any of its subclasses);
\item leaving all the statements in $I'_0$ if these conditions do not
  hold.
\end{itemize}
Basically, this single-field optimization identifies cases where the
only cycles or reachability paths starting from $v$ must forcefully
traverse $f$, either because $f$ is the only field, or because no
other field makes such cycles or paths possible.  It must be pointed
out that this optimization relies on information about classes and
fields which can be obtained statically by code inspection, and was
not included in the original analysis of Ghiya and
Hendren~\cite{DBLP:conf/popl/GhiyaH96}.
The sets $I_r$ and $I_c$ capture the effect of executing
$v.f\assign\res$ on $I'$.  Moreover, $I'_s$ contains sharing, purity
and aliasing information after evaluating $\mathit{exp}$.

The following reachability statements are added: for any $w_1$ which
might either alias with $v$ or reach $v$ (formalized as
$(\ALIAS{w_1}{v} \in  I'_s) \vee (\REACHES{w_1}{v}  \in  I')$), and
any $w_2$ aliasing with $\res$ or reachable from it (formalized as
$(\ALIAS{\res}{w_2} \in  I'_s) \vee (\REACHES{\res}{w_2} \in  I')$),
the statement $\REACHES{w_1}{w_2}$ is added since the new path created
by the update implies that $w_1$ can reach $w_2$.  This accounts for
all possible paths which can be created by adding a direct link from
$v$ to $\res$ through $f$.

New cyclicity statements are contained in $I_c$.  There are three
possible scenarios where $v$ might become cyclic:
\begin{itemize}
\item $\res$ reaches $v$, so that a cycle from $v$ to itself is
  created;
\item $\res$ aliases with $v$, so that $v$ reaches itself with a path
  of length $1$ (e.g., the command \lstinline!y.f:=y!); or
\item $\res$ is cyclic, so that $v$ becomes indirectly cyclic.
\end{itemize}
Whenever one of these scenarios occurs (formalized as
$(\REACHES{\res}{v} \in  I') \vee (\ALIAS{\res}{v} \in  I'_s) \vee
(\CYCLE{\res} \in  I')$), any variable $w$ aliasing with $v$ or
reaching it (formalized as $(\ALIAS{w}{v}  \in  I'_s) \vee
(\REACHES{w}{v}  \in  I')$) has to be considered as possibly cyclic.

\begin{example}
  \label{ex:field-up}
  Consider line 20 in Figure~\ref{fig:run-ex-code}.  The abstract
  value before such a line, produced at line 17, is
  $I_{17}=\{\REACHES{\this}{c} , \REACHES{\this}{p}, \REACHES{p}{c} ,
  \REACHES{n}{c}\}$.  The evaluation of
  $\ASEMANTICS{\typenv}{\absinterp}{\mbox{\lstinline!p.next:=n!}}(I_{17})$
  at line 20 adds a new statement $\REACHES{p}{n}$, as expected.
  Moreover, it also adds $\REACHES{\this}{n}$ since \codethis was
  reaching \pp, and both $\REACHES{p}{c}$ and $\REACHES{\this}{c}$
  (which, however, were already contained in $I_{17}$) since \nn was
  reaching \cc.
\end{example}

\subsubsection{Conditions, loops, composition, and return command}

Rules $(3_c){:}~\ASEMANTICS{\typenv}{\absinterp}{
  \ifte~\mathit{exp}~\iftethen~\mathit{com}_1~\ifteelse~\mathit{com}_2}$,
$(4_c){:}~\ASEMANTICS{\typenv}{\absinterp}{\while~\mathit{exp}~\whilebody~\mathit{com}}$,
and $(6_c){:}~\ASEMANTICS{\typenv}{\absinterp}{\mathit{com}_1;
  \mathit{com}_2}$ are quite straightforward and correspond,
respectively, to the \emph{if} conditional, the \emph{while} loop, and
\emph{command composition}.  Finally,
rule~$(5_c){:}~\ASEMANTICS{\typenv}{\absinterp}{\return~\mathit{exp}}$
corresponds to the \emph{return} command, and behaves, as expected,
like the execution of $\out\mbox{\lstinline!:=!}\mathit{exp}$.

\subsubsection{Method calls}
\label{sec:methodCalls}

Rule $(7_e){:}~\EXPASEMANTICS{\typenv}{\absinterp}{v_0.m(v_1,..,v_n)}$
propagates the effect of a method call to the calling context, as
follows:

\begin{enumerate}
\item the abstract state $I$ is projected on the actual parameters
  $\bar{v}$, thus obtaining $I_0$; this is needed since the denotation
  of the callee is given in terms of its parameters;
\item the denotation of each method $\methodsig{m}$ which can be
  possibly called at runtime is taken from the current interpretation,
  namely, $\absinterp(\methodsig{m})$, and applied to
  $I_0[\bar{v}/\inp{\methodsig{m}}]$, which is the result of renaming
  the actual parameters $\bar{v}$ to the formal parameters
  $\inp{\methodsig{m}}$ in $I_0$;
\item formal parameters are renamed back to the actual parameters
  (plus $\out$ and $\res$) in the resulting state
  $\absinterp(\methodsig{m})(I_0[\bar{v}/\inp{\methodsig{m}}])$, and
  the states obtained from all possible signatures are merged into
  $I_m$.
\end{enumerate}

Step 2 takes more than one method into account because, in an
Object-Oriented language with inheritance, it is in general not
possible to decide, at compile-time, which method instance (among
various method declarations whose signature is compatible with the
type of the actual parameters and the expected return value) will be
actually invoked after calling the function $\lookup$ (Section
\ref{sec:language}).  Therefore, the abstract semantics takes,
conservatively, the union of all of them.

In the definition, $I'_s$ is a safe approximation of the
sharing among actual parameters, and $I''_s$ safely approximates the
sharing and purity information after the method call.  The definitions
of $I_1$, $I_2$, $I_3$, and $I_4$ account for the propagation of the
effects of the method execution in the calling context:

\begin{itemize}
\item $I_1$ states that, if the call creates reachability from $v_i$
  to $v_j$, then any $w_1$ sharing with $v_i$ before the call might
  reach any $w_2$ which is reachable from $v_j$ or aliasing with
  $v_j$.  Note that adding these statements is necessary only if $v_i$
  is updated in the body of some $\methodsig{m}$ (this information is
  conservatively represented in $I''_s$, so that the condition
  $\PURE{v}_i  \in  I''_s$ must be checked): otherwise, it is
  guaranteed that no path from $w_1$ to $w_2$ will be created during
  the call.
\item $I_2$ states that, if the call makes $v_i$ share with $v_j$,
  then any $w_1$ sharing with $v_i$ might reach any $w_2$ reachable
  from $v_j$.  Again, this is required only if $v_i$ is updated in the
  body of any $\methodsig{m}$.
\item $I_3$ contains the information about any variable $v$ aliasing
  with $\res$, which is cloned for $\res$.
\item $I_4$ will include the possible cyclicity of anything sharing
  with an argument which might become cyclic.
\end{itemize}
The final result of processing a method call is the union $I\cup
I_m\cup I_3\cup I_4$.

\begin{figure}[t]
\begin{center}
\begin{minipage}[t]{5.5cm}
\begin{lstlisting}
   Node f(Node a,Node b,Node c) {
     a.next:=b;
     c.next:=this;
     return b.g(c); 
   }
\end{lstlisting}
\end{minipage}
~~~~~~~
\begin{minipage}[t]{3.5cm}
\begin{lstlisting}
   Node g(Node y) {
     this.next:=y;
     return this; 
   }
\end{lstlisting}
\end{minipage}

~\\~\\

\noindent
\begin{minipage}[t]{5.5cm}
\begin{lstlisting}
   Node h(Node y) {
     this.next:=y;
     y:=null;
     return this; 
   }
\end{lstlisting}
\end{minipage}
~~~~~~~
\begin{minipage}[t]{3.5cm}
\begin{lstlisting}
   Node k(Node y) {
     u:=y;
     this.next:=y;
     y:=null;
     return this; 
   }
\end{lstlisting}
\end{minipage}
\end{center}

\caption{Some more examples}
\label{fig:more-ex}
\end{figure}

\begin{example}
  \label{ex:method-call}
  Consider methods \lstinline!f! and \lstinline!g! of
  Figure~\ref{fig:more-ex}, and assume that both are defined in the
  class \lstinline!Node!.  
  Let $\absden$ be a denotation for
  \lstinline!g! such that
  $\absden(\emptyset)=\{\REACHES{\this}{y},\REACHES{out}{y}\}$.  This
  example shows how an abstract state $\emptyset$ is transformed by
  executing the code of \lstinline!f!.
  The first two commands in \lstinline!f! transform $\emptyset$ into
  $I=\{\REACHES{a}{b},\REACHES{c}{\this}\}$.
  Then, the denotation of \lstinline!g! is plugged into the calling
  context, as follows:
  \begin{enumerate}
  \item $I$ is projected on $\{b,c\}$, obtaining $I_0=\emptyset$;
  \item $\absden(\emptyset)$ is renamed such that $\this$, $y$, and
    $\out$ are renamed to, respectively, $b$, $c$, and $\res$, and
    $I_m=\{\REACHES{b}{c},\REACHES{\res}{c}\}$ is obtained;
  \item $\REACHES{a}{\this}$ is added to $I_1$ since
    $(\REACHES{b}{c}\in I_m)\wedge(\SHARE{b}{a}\in
    I'_s)\wedge(\REACHES{c}{\this} \in I)$ is true; similarly,
    $\REACHES{b}{\this}$, $\REACHES{a}{c}$ and $\REACHES{a}{\res}$ are
    also added to $I_1$;
  \item no new statements have to be added because of $I_2$ or $I_3$;
  \item $I_4$ is empty since nothing becomes cyclic in \lstinline!g!;
  \item finally, the denotation of \return renames $\res$ to $\out$ in
    $I\cup I_m\cup I_1\cup I_4$, and obtains
    $\{\REACHES{a}{b},\REACHES{c}{\this},\REACHES{b}{c},\REACHES{\out}{c},
    \REACHES{a}{\this},\REACHES{a}{c},\REACHES{b}{\this},
    \REACHES{a}{\out}\}$.
  \end{enumerate}
\end{example}

\noindent
Next, the inference of a denotation for a method $\methodsig{m}$ is
shown, which uses the denotation
$\ASEMANTICS{\typenv}{\absinterp}{\body{\methodsig{m}}}$ of its code.
Example~\ref{ex:need-for-shallow} introduces the problems to be faced
when trying to define a method denotation, and a solution is discussed
below.

\begin{example}
  \label{ex:need-for-shallow}
  In Example~\ref{ex:method-call}, when analyzing \lstinline!b.g(c)!,
  the existence of a denotation $\absden$ for \lstinline!g!  such that
  $\absden(\emptyset)=\{\REACHES{\this}{y},\REACHES{\out}{y}\}$ was
  assumed.
  Intuitively, this $\absden(\emptyset)$ could be computed using
  $\ASEMANTICS{\typenv}{\absinterp}{\body{\methodsig{g}}}$, as
  follows: the first command in \lstinline!g! adds
  $\REACHES{\this}{y}$, and the second one adds $\REACHES{\out}{y}$,
  which results in the desired abstract state
  $\{\REACHES{\this}{y},\REACHES{\out}{y}\}$.
  After this result, one might think that
  $\ASEMANTICS{\typenv}{\absinterp}{\body{\methodsig{m}}}(I)$ is
  always the good way to compute $\absden(I)$, as just done.  Yet, in
  general, this is not correct.
  For example, suppose the call $\mbox{\lstinline!b.g(c)!}$ is
  replaced by \lstinline!b.h(c)! (which is defined in Figure~\ref{fig:more-ex} also).
  The effect of this call should be
  the same as \lstinline!b.g(c)!, since both methods make $b$ reach
  $c$ and $b$ reach the return value.
  However, computing
  $\ASEMANTICS{\typenv}{\absinterp}{\body{\methodsig{h}}}(\emptyset)$
  has a different result: the first instruction adds
  $\REACHES{\this}{y}$, but the second one removes it since the value
  of $y$ is overwritten, and the third does not add anything.
  Therefore,
  $\ASEMANTICS{\typenv}{\absinterp}{\body{\methodsig{h}}}(\emptyset)=\emptyset$,
  which is not sound to use as the result of $\absden(\emptyset)$.
\end{example}

\noindent
The problem in Example~\ref{ex:need-for-shallow} comes from the
call-by-value passing style for parameters, where, if the formal
parameters are modified in the method, then the final abstract state
does not describe the actual parameters anymore.
This is why the expected reachability information is obtained for
\lstinline!f! (since it does not modify \lstinline!y!), while it is
not in the case of \lstinline!h! (since \lstinline!y! is modified in
the body).
A common solution to this problem is to mimic actual parameters by
\emph{shallow variables} or \emph{ghost variables}, i.e., new
auxiliary variables which are initialized when entering the method to
the same values as the parameters, but are never modified in the body.

\begin{example} Consider methods \lstinline!h! and \lstinline!k! in
  Figure~\ref{fig:more-ex}.
  Method \lstinline!k!  is the result of instrumenting
  \lstinline!h!  with a shallow variable \lstinline!u!, mimicking
  \lstinline!y!.
  It is easy to verify that
  $\ASEMANTICS{\typenv}{\absinterp}{\body{\methodsig{k}}}(\emptyset)$
  comes to be $\{\REACHES{\this}{u}, \REACHES{\out}{u}\}$, which
  includes the desired reachability information.
\end{example}

  The following definition defines the abstract denotational semantics
  of a program $P$ as the least fixpoint of an (abstract) transformer
  of interpretations.  Variables $\bar{u}$ play the role of shallow
  variables.  Note that shallow variables appear at the level of the
  semantics, rather than by transforming the program.

\begin{definition}
  \label{def:abs-den-semantics}
  The \emph{abstract denotational semantics} of a program $P$ is the
  \emph{lfp} of the transformer
  \[
  \abstp{P}(\absinterp) = \{ \methodsig{m}\mapsto\lambda
  I \in \domrc{\inp{\methodsig{m}}} (\project
  {\ASEMANTICS{\typenv}{\absinterp}{\body{\methodsig{m}}}(I\cup
    I[\bar{w}/\bar{u}])} {X})[\bar{u}/\bar{w}] ~|~
  \begin{array}{l}
    \methodsig{m}\in P\\
    
  \end{array}
  \}
  \]
  
  \noindent
  where $\inp{\methodsig{m}} = \{\this,w_1,\ldots,w_n\}$, and
  $\bar{u}$ is a variable set $\{u_1,\ldots,u_n\}$ such that
  $\bar{u}\cap \scope{\methodsig{m}}=\emptyset$; moreover,
  $\dom(\tau)= \locals{\methodsig{m}}\cup \bar{u}$, and
  $X=\dom(\tau){\setminus}(\bar{u}\cup \{\this,\out\})$.
\end{definition}

\noindent
The definition is explained in the following.  The operator
$\abstp{P}$ transforms the interpretation $\absinterp$ by assigning a
new denotation for each method $\methodsig{m}\in P$, using those in
$\absinterp$.
The new denotation for $\methodsig{m}$ maps a given input abstract
state $I\in\domrc{\inp{\methodsig{m}}}$ to an output state abstract
from $\domrc{\inp{\methodsig{m}}\cup\{\out\}}$, as follows:
\begin{enumerate}
\item it obtains an abstract state $I_0=I \cup I[\bar{w}/\bar{u}]$ in
  which the parameters $\bar{w}$ are cloned into the shallow variables
  $\bar{u}$;
\item it applies the denotation of the code of $\methodsig{m}$ on
  $I_0$, obtaining
  $I_1=\ASEMANTICS{\typenv}{\absinterp}{\body{\methodsig{m}}}(I_0)$;
\item all variables but $\bar{u}\cup\{\this,\out\}$ are eliminated
  from $I_1$ (using $\exists X$); and
\item shallow variables $\bar{u}$ are finally renamed back to
  $\bar{w}$.
\end{enumerate}

\noindent
Soundness is addressed in Section~\ref{sec:soundness}, next we see
some examples.


\begin{example}
\label{ex:mirror}
Consider the following method 
  
\medskip
\begin{lstlisting}
  int mirror(Tree t) {
    Tree l,r;
  
    if (t=null) then { 
      return 0;
    } else {
      l:=t.left;
      r:=t.right;
      t.left:=r;
      t.right:=l;
      return 1+mirror(l)+mirror(r); 
    } 
  }
\end{lstlisting}
  
  \noindent
  and suppose that class \lstinline!Tree! implements binary trees in
  the standard way, with fields \lstinline!left! and
  \lstinline!right!.
  The call \lstinline!mirror(t)! exchanges the values of
  \lstinline!left! and \lstinline!right! of each node in
  \lstinline!t!, and returns the number of nodes in the tree.
  An initial state $\emptyset$ is transformed by \lstinline!mirror! as
  follows.
  Suppose that the current interpretation $\absinterp$ is such that
  $\absinterp(\methodsig{mirror})=\absden$, and
  $\absden(\emptyset)=\emptyset$.
  The first branch of the \emph{if} (when \lstinline!t! is \nil) does
  not change the initial denotation; on the other hand, when
  \lstinline!t! is different from \nil, line 7 adds $\REACHES{t}{l}$;
  line 8 adds $\REACHES{t}{r}$; line 9 adds again $\REACHES{t}{r}$;
  and line 10 adds again $\REACHES{t}{l}$.  Recursive calls
  \lstinline!mirror(l)!  and \lstinline!mirror(r)! do not add any
  statement since $\absden(\emptyset)=\emptyset$.  Finally, \return
  adds nothing.
  Projecting $\{\REACHES{t}{l},\REACHES{t}{r}\}$ on $t$ and $\out$
  results in $\emptyset$, so that $\absden(\emptyset)$ does not
  change, and there is no need for another iteration.
  It can be concluded that, as expected, mirroring the tree does not
  make it cyclic.
\end{example}

\begin{example}
\label{ex:connect}
Consider the following method 
  
\medskip
\begin{lstlisting}
  Node connect() {
    Node curr;
    
    curr=this;
    while (curr.next!=null) {
      curr:=curr.next; 
    }
    curr.next:=this;
    return curr; 
  }
\end{lstlisting}
 
  \noindent
  and assume it is defined in the class \lstinline!Node!. A call
  \lstinline!l.connect()! with \lstinline!l! acyclic makes the last
  element of \lstinline!l! point to \lstinline!l!, so that it becomes
  cyclic.  It also returns a reference to the last element in the
  list.
  An initial state $\emptyset$ is transformed by \lstinline!connect!
  as follows.  Line 4 does not add any statements, while line 6 in the
  loop adds $\REACHES{\this}{\mathit{curr}}$.  Another iteration of
  the loop does not change anything, so that the loop is exited with
  $\{\REACHES{\this}{\mathit{curr}}\}$.
  Since $\codethis$ is now reaching \lstinline!curr!, line 8 adds $\{
  \REACHES{\mathit{curr}}{\this}$,
  $\REACHES{\mathit{curr}}{\mathit{curr}}, \REACHES{\this}{\this}\}$,
  and $\{ \CYCLE{\mathit{curr}}, \CYCLE{\this} \}$.
  Finally, line 9 clones $\mathit{curr}$ to $\out$.
  In conclusion, the analysis correctly infers that
  \lstinline!l.connect()!  makes \lstinline!l! and the return value
  cyclic.
\end{example}

\subsection{Soundness}
\label{sec:soundness}

This section present the soundness theorem: the abstract state
obtained by applying the abstract semantics to a method in a given
input abstract state is a \emph{correct representation} of (i.e., its
concretization contains) the concrete state obtained by executing the
method in any input concrete state which is correctly represented by
such input abstract state.  The proof of the theorem can be found in
\ref{sec:proofs:4}.


\begin{theorem}[Soundness]
  \label{th:soundness}
  Let $P$ be a program, and $\interp$ and $\absinterp$ be,
  respectively, its concrete and abstract semantics according to
  Definitions~\ref{def:conc-den-semantics} and
  \ref{def:abs-den-semantics}.  Moreover, let $\methodsig{m}$ be a
  method in $P$, and let $\den=\interp(\methodsig{m})$ and
  $\absden=\absinterp(\methodsig{m})$.
  It holds that, for all $\statesym_1\in\states{\inp{m}}$,
  \[ \statesym_2=\den(\statesym_1) \qquad \Rightarrow \qquad 
  \state{\statef{\statesym}_1[\out \mapsto
      \statef{\statesym}_2(\out)]}{\statem{\statesym}_2} \in
  \gammarc(\absden(\alpharc(\{\statesym_1\}))) \]
\end{theorem}

\subsection{Completeness and optimality}
\label{sec:completeness}

\emph{Completeness} \cite{GQ01} is a well-known notion in Abstract
Interpretation, and corresponds to require that no loss of precision
is introduced by computing an abstract semantic function on abstract
states with respect to approximating the same (concrete) computation
on concrete states.
An abstract domain $A$ (with abstract function $\alpha$ and
concretization function $\gamma$) and an abstract function $f^{\#}$
over it are \emph{backward-complete} for the concrete function $f$ if
and only if, for every concrete input $\statesym$, the abstraction
$\alpha(f(\statesym))$ of a concrete computation is equal to the abstract
computation $f^{\#}(\alpha(\statesym))$. 
This property guarantees that 
$\alpha(\mathit{lfp}(f))=\mathit{lfp}(f^{\#})$.

By \emph{optimality} we refer to the fact that the abstract function
under study is the \emph{best correct approximation} of the concrete
function with respect to the associated abstraction: for every $I$,
$f^{\#}(I)$ must be equal to $\alpha(f(\gamma(I)))$.


For the sake of the following discussion, the abstract semantics
$\ASEMANTICS{\typenv}{\absinterp}{\_}$ (a similar discussion holds for
$\EXPASEMANTICS{\typenv}{\absinterp}{\_}$) is supposed to use, for
collecting sharing, aliasing and purity information, the \emph{best
  correct approximation}
$\mathcal{S}_{\typenv}^{\absinterp}{\left\llbracket{\_}\right\rrbracket}$
of $\cinter{\typenv}{\interp}{\_}$ with respect to
$\cI_{s}^{\typenv}$: for every command $\mathit{com}$ and abstract
value $I \in \cI_{s}^{\typenv}$,
$\mathcal{S}_{\typenv}^{\absinterp}{\left\llbracket{\mathit{com}}\right\rrbracket}(I)$
is defined as
$\alpha_s(\cinter{\typenv}{\interp}{\mathit{com}}(\gamma_s(I)))$.  To
introduce the abstract semantics over this domain is necessary in
order to be able to properly talk about completeness and optimality of
the reachability and cyclicity analysis, as it will be clear in the
following.

\paragraph{Backward completeness}

The present analysis is not backward-complete.  In the following, the
abstract domain under study will be $\domrc{\typenv} \sqcap
\cI_{s}^{\typenv}$ (i.e., sharing, aliasing and purity are included).
Consider the state $\statesym$ obtained by executing the following
statements, starting from a heap where all variables are \nil: the
final result of the execution is the heap shown in the picture.

\medskip
\begin{minipage}{6cm}
  \begin{lstlisting}
    y:=new C;
    z:=new C;
    y.f:=new C;
    z.f:=y.f;
    y.g:=z;
  \end{lstlisting}
\end{minipage}
\begin{minipage}{6cm}
\begin{tikzpicture}
  \tikzstyle{node}=[draw,circle,minimum size=4mm]
  \tikzstyle{ordl}=[draw,minimum size=4mm]
  \tikzstyle{myvar}=[minimum size=2mm]
  \tikzstyle{myderef}=[->,double]
  \node[ordl]  (yloc) at (0  ,0   ) {};
  \node[ordl]  (zloc) at (2  ,0   ) {};
  \node[ordl]  (loc)  at (1  ,-1   ) {};
  \node[myvar] (y)    at (-1,0.3) {\lstinline[basicstyle=\scriptsize\sffamily]!y!};
  \node[myvar] (z)    at ( 3,0.3) {\lstinline[basicstyle=\scriptsize\sffamily]!z!};

  \draw[->] (y) -- (yloc);
  \draw[->] (z) -- (zloc);
  \draw[myderef] (yloc) -- node[auto] {\lstinline[basicstyle=\scriptsize\sffamily]!f!} (loc);
  \draw[myderef] (zloc) -- node[auto] {\lstinline[basicstyle=\scriptsize\sffamily]!f!} (loc);
  \draw[myderef] (yloc) -- node[auto] {\lstinline[basicstyle=\scriptsize\sffamily]!g!} (zloc);
\end{tikzpicture}

\end{minipage}

\noindent
After this code fragment, \yy and \zz share because they reach a
common location, and \yy is reaching \zz.  Then, the most precise
approximation of the resulting concrete state $\statesym$ is $I = \{
\SHARE{y}{y}, \SHARE{z}{z}, \SHARE{y}{z}, \ALIAS{y}{y}, \ALIAS{z}{z},
\REACHES{y}{z} \}$\footnote{The notation $\SHARE{\_}{\_}$ and
  $\ALIAS{\_}{\_}$ is used in the beginning of Section
  \ref{sec:abstractSemantics}}.  Suppose that the statement

\medskip
\begin{minipage}{6cm}
  \begin{lstlisting}[firstnumber=6]
    x:=y.f;
  \end{lstlisting}
\end{minipage}

\noindent
is executed afterward, giving the concrete state $\statesym'$: in this
case, the concrete function $f$ under study is the semantic of this
statement, namely,
$\cinter{\typenv}{\absinterp}{\mbox{\lstinline!x:=y.f!}}$, and the
state $\statesym'$ corresponds to
$\cinter{\typenv}{\absinterp}{\mbox{\lstinline!x:=y.f!}}(\statesym)$.
Now, the abstraction of $\statesym'$ with respect to $\domrc{\typenv}
\sqcap \cI_{s}^{\typenv}$ is \[I' = \{ \SHARE{x}{x}, \SHARE{y}{y},
\SHARE{z}{z}, \SHARE{x}{y}, \SHARE{x}{z}, \SHARE{y}{z}, \ALIAS{x}{x},
\ALIAS{y}{y}, \ALIAS{z}{z}, \REACHES{y}{x}, \REACHES{y}{z},
\REACHES{z}{x} \}\] which correctly represents the sharing between the
three variables, and the fact that \xx points exactly to the location
which is reached by both \yy and \zz.  On the other hand, computing
the result of the abstract semantics $f^{\#}$ (i.e., the present
analysis
$\ASEMANTICS{\typenv}{\absinterp}{\mbox{\lstinline!x:=y.f!}}$) on the
input abstract state $I$ gives the state \[\begin{array}{rl}I'' = & \{
  \SHARE{x}{x}, \SHARE{y}{y}, \SHARE{z}{z}, \SHARE{x}{y},
  \SHARE{x}{z}, \SHARE{y}{z}, \\ & \ALIAS{x}{x}, \ALIAS{y}{y},
  \ALIAS{z}{z}, \ALIAS{x}{y}, \ALIAS{x}{z}, \ALIAS{y}{z},
  \REACHES{y}{x}, \REACHES{y}{z}, \REACHES{z}{x}, \REACHES{x}{z}
  \}\end{array}\] The reachability statement $\REACHES{x}{z}$ is added
because the analysis admits that, since \yy is said to reach \zz, the
location pointed to by \xx could be exactly on the path from \yy to
\zz.  Because of the difference between $I'$ and $I''$, this
counterexample is enough to prove the lack of backward completeness.




\paragraph{Optimality}

This section argues that two important abstract state transformers
included in the abstract semantics are optimal.  The considered
transformers are $f_1^\# = \EXPASEMANTICS{\typenv}{\absinterp}{v.f}$
for field access, which is optimal with respect to $f_1 =
\einter{\typenv}{\interp}{v.f}$, and $f^\#_2 =
\ASEMANTICS{\typenv}{\absinterp}{v.f\assign \res}$ for field updates,
which is optimal with respect to $f_2 =
\cinter{\typenv}{\interp}{v.f\assign\res}$.  The use of $\res$ means
that the state transformers account for the field update \emph{after}
the expression $\mathit{exp}$ has been evaluated.  In other words,
$f_2$ will be applied to the concrete state resulting from evaluating
$\mathit{exp}$, and $f^\#_2$ will be applied to the abstract state
$I'_0$ described in Figure \ref{fig:abs-sem-com}.  In order to avoid
confusion with names, let $J$ be the abstract value which is given as
input to the abstract state transformer, and let $J_1$ the
corresponding output; therefore, $J'$ and similar names will play the
same role as $I'$ and similar names in Figure \ref{fig:abs-sem-com}.

Again, the abstract domain includes sharing, aliasing and purity, so
that the concretization and abstraction functions $\gamma$ and
$\alpha$ are the ones which are induced by the reduced product
$\domrc{\typenv} \sqcap \cI_{s}^{\typenv}$ in the standard way.  This
means that optimality is proven under the assumption that the abstract
operators of sharing, aliasing and purity are also optimal.  It is
assumed that an abstract value contains sharing, aliasing and purity
information, together with reachability and cyclicity, and that it
will be clear from the context how to refer to each part.

By soundness, the non-strict inequalities
$\EXPASEMANTICS{\typenv}{\absinterp}{v.f}(J) \supseteq
\alpha(\einter{\typenv}{\interp}{v.f}(\gamma(J)))$ and
$\ASEMANTICS{\typenv}{\absinterp}{v.f\assign \res}(J) \supseteq
\alpha(\cinter{\typenv}{\interp}{v.f\assign \res}(\gamma(J)))$ already
hold, where set inclusion is the partial order on $\domrc{\typenv}
\sqcap \cI_{s}^{\typenv}$.  Therefore, to prove this claim amounts to
demonstrate the other direction of the inclusion, i.e., that, for
every reachability or cyclicity statement $\mathit{st}$ contained in
$J$, there is a concrete state $\statesym \in \gamma(J)$ such that
$\statesym_1 = \cinter{\typenv}{\interp}{v.f\assign \res}(\statesym)$
(the case of $\einter{\typenv}{\interp}{v.f}(\statesym)$ is similar)
is a concrete state whose abstraction $\alpha(\{\statesym_1\})$
contains $\mathit{st}$.  In other words, $\statesym_1$ is a state
where the may-information represented by $\mathit{st}$ is
\emph{actually} happening (for example, if $\mathit{st}$ is some
$\REACHES{v}{w}$, then there must \emph{actually} be a path in the
heap from $v$ to $w$ in $\statesym_1$), so that the abstraction of
$\statesym_1$ will forcefully contain such a statement.  In the proof,
this idea of ``a statement $\mathit{st}$ actually happening in a state
$\statesym$'' will be phrased as ``$\statesym$ \emph{justifies}
$\mathit{st}$''.

\begin{itemize}
\item Case $f^\#_1$: the output abstract state $J_1$ is basically the
  union of four sets: (a) $J$; (b) $J[v/\res]$; (c) $\{
  \REACHES{w}{\res}~|~\SHARE{w}{v}{\in} J_s \}$; and (d) $\{
  \REACHES{\res}{\res}~|~\CYCLE{v} \in J \}$.  For every one of them
  it is necessary to prove that, for every statement $\mathit{st}$
  contained in it, there exists at least one concrete input state
  $\statesym$ such that the corresponding output state $\statesym_1 =
  f_1(\statesym)$ justifies $\mathit{st}$.
  \begin{itemize}
  \item[(a)] Clearly, every statement $\mathit{st}$ which was already
    in $J$, and is therefore maintained in $J_1$, is justified by the
    fact that the structure of the heap does not change when
    evaluating the expression: by hypothesis, there was already a state
    $\statesym$ justifying $\mathit{st}$, and the corresponding output
    $\statesym_1$ still justifies such statement.
    
  \item[(b)] In this case, relevant statements in $J$ can be of four
    kinds (other statements which do not involve $v$ are not
    relevant), and we need to prove that the corresponding statements
    in $J[v/\res]$ (where $v$ is replaced by $\res$) are justified.
    \begin{itemize}
    \item[$\REACHES{v}{w}$]: In this case, there certainly exists
      $\statesym$ in the concretization of $J$ such that $v$ actually
      reaches $w$ in at least two steps, and the first step goes
      through $f$; then, the location pointed to by the expression
      actually reaches $w$ (in fact, it is on the path from $v$ to
      $w$), so that $\statesym_1$ justifies the statement
      $\REACHES{\res}{w}$ contained in $J[v/\res]$, corresponding to
      $\REACHES{v}{w}$;
    \item[$\REACHES{w}{v}$]: This case is easy since there exists
      $\statesym$ such that $w$ actually reaches $v$, and it is
      straightforward to see that $\res$ will be actually reached by
      $w$ in $\statesym_1$ (transitivity of reachability at the
      concrete level), thus justifying the corresponding statement
      $\REACHES{w}{\res}$ in $J[v/\res]$;
    \item[$\REACHES{v}{v}$]: This case is also easy because there
      certainly exists $\statesym$ such that $v$ is cyclic, and the
      first step of the cycle when starting from $v$ goes through $f$;
      this means that $v.f$ is still in the cycle, and the location
      pointed to by the expression reaches itself, thus justifying the
      corresponding statement $\REACHES{\res}{\res}$ in $J[v/\res]$;
    \item[$\CYCLE{v}$]: This case is similar to the previous one.
    \end{itemize}
    
  \item[(c)] In this case, every $\REACHES{w}{\res}$ must be
    justified, provided there is sharing (this is a case where it
    becomes clear that sharing must also be considered) between $v$
    and $w$ in the input state.  It is enough to take the same (up to
    variable renaming) concrete state used in the discussion about
    backward completeness, where $v$ and $w$ both reach (in one step,
    and through $f$) the same location in the heap: the location
    pointed to by $\res$ in the output state comes to be actually
    reached by $w$, thus justifying the statement.

  \item[(d)] The last case is easy because it is enough to find some
    $\statesym$ where $v$ is cyclic (but not necessarily reaching
    itself), and the location pointed to by $v.f$ reaches itself.
  \end{itemize}

\item Case $f^\#_2$: The first issue here is to note that optimality
  requires the single-field optimization discussed in Section
  \ref{sec:fieldUpdate}, where $J_0$ is strictly smaller than $J'_0$
  whenever it can be guaranteed that all the relevant reachability or
  cyclicity paths have been broken by updating $v.f$.  In fact,
  consider the case where this optimization is not performed (i.e.,
  $J' = J'_0$).  The following piece of code
  \begin{lstlisting}
      x := new C();
      x.f := x;
      x.f := null;
  \end{lstlisting}
  shows the lack of optimality under the condition that \lstinline!f!
  is the only field of \lstinline!C!.  In fact, let the abstract value
  $J$ before line 3 be $\{ \REACHES{\xx}{\xx}, \CYCLE{\xx} \}$ as it
  would be obtained by the analysis, so that $\gamma(J)$ contains all
  the states where \xx is cyclic and reaches itself.  However, the
  abstract semantics without the optimization would generate the same
  abstract value $\{ \REACHES{\xx}{\xx}, \CYCLE{\xx} \}$ as the final
  value.  This is not optimal since any concrete state after executing
  this code would have \xx acyclic and not self-reaching, so that its
  abstraction would be $\{\}$ (in other words, none of the statements
  would be justified).  On the other hand, the aforementioned
  optimization removes these statements from $J'_0$, so that $J'$ is
  empty, thus achieving, in the end, optimality.

  In the definition depicted in Figure \ref{fig:abs-sem-com}, the
  output abstract state $J_1$ consists of two more parts: (a) the one
  coming from $J_r = \{ \REACHES{w_1}{w_2} ~|~ ((\ALIAS{w_1}{v}{\in}
  J'_s) \vee (\REACHES{w_1}{v} \in J'))\wedge ((\ALIAS{\res}{w_2}{\in}
  J'_s) \vee (\REACHES{\res}{w_2} \in J')) \}$; and (b) the one coming
  from $J_c = \{ \CYCLE{w}~|~ ((\REACHES{\res}{v} \in J') \vee
  (\ALIAS{\res}{v}{\in} J'_s) \vee (\CYCLE{\res} \in J'))\wedge
  ((\ALIAS{w}{v}{\in} J'_s) \vee (\REACHES{w}{v} \in J')) \}$.
  \begin{itemize}
  \item[(a)] In order to justify a statement $\REACHES{w_1}{w_2}$, it
    is enough to take a concrete state $\statesym \in \gamma_s(J)$
    (which clearly exists) where $w_1$ is actually reaching $v$, and
    the location pointed to by the result of the expression is
    actually reaching $w_2$.  In this case, the field update will
    create a path from $w_1$ to $w_2$ in $\statesym_1$, so that the
    statement is justified.
    
  \item[(b)] A statement $\CYCLE{w}$ can be easily justified by taking
    $\statesym$ such that the result of the expression points to an
    actually cyclic data structure, and $w$ actually reaches $v$.
    Then, the newly created path will make $w$ cyclic.
  \end{itemize}

  The final elimination of $\res$ is not problematic.
  
\end{itemize}

%

%% file: abs-semantics.tex

\begin{figure}[t]
    \begin{minipage}{11.7cm}
\[
      \small
      \begin{array}{@{}l@{~}rl@{}}
        %
        (1_e) & \EXPASEMANTICS{\typenv}{\absinterp}{n}(I) =& I \\
        (2_e) & \EXPASEMANTICS{\typenv}{\absinterp}{\nil}(I) =& I \\
        (3_e) & \EXPASEMANTICS{\typenv}{\absinterp}{\newk{\class}}(I) =& I
        \\
        (4_e) & \EXPASEMANTICS{\typenv}{\absinterp}{v}(I)    =& 
        \mbox{if $\typenv(v){=}\integer$ then $I$ else $I\cup
          I[v/\res]$}
        \\
        (5_e) & \EXPASEMANTICS{\typenv}{\absinterp}{v.f}(I)  =& 
        \mbox{if $f$ has type \integer then $I$ else $I\cup I'$ where}\\
        && ~I' {=} I[v/\res] \cup \{ \REACHES{w}{\res}|\SHARE{w}{v}{\in}I_s \}\cup 
        \{ \REACHES{\res}{\res}~|~\CYCLE{v} \in I \}
        \\
        (6_e) & \EXPASEMANTICS{\typenv}{\absinterp}{\mathit{exp}_1 \oplus
          \mathit{exp}_2}(I) =& \exists
        \rho.\EXPASEMANTICS{\typenv}{\absinterp}{\mathit{exp}_2}(\exists
        \rho.\EXPASEMANTICS{\typenv}{\absinterp}{\mathit{exp}_1}(I)) \\
        (7_e)
        &\EXPASEMANTICS{\typenv}{\absinterp}{v_0.m(v_1,..,v_n)}(I)
        =&  I \cup I_m \cup I_3 \cup I_4~\mbox{where} \\ 
        &\multicolumn{2}{l}{~~~~~~~\bar{v}{=}\{v_0,..,v_n\}} \\ &\multicolumn{2}{l}{~~~~~~~I_0 {=}
          \project{I}{(\typenv{\setminus}\bar{v})}} \\
        &\multicolumn{2}{l}{~~~~~~~
          I_m = \cup~\{ 
          ~(\absinterp(\methodsig{m})(I_0[\bar{v}/\inp{\methodsig{m}}]))[\inp{\methodsig{m}}/\bar{v},\out/\res]~
          ~|~ \methodsig{m} \mbox{ might be called here\footnote{See
              Section \ref{sec:methodCalls} } }\} }
        \\
        &\multicolumn{2}{l}{~~~~~~~I'_s=\{\SHARE{v_i}{v_j} ~|~
          v_i,v_j\in\bar{v} \mbox{ and } \SHARE{v_i}{v_j}\in I_s \}
          \cup \{ \PURE{v}|v \in \bar{v}  \mbox{ and } \PURE{v} \in
          I_s \}
        }\\
        &\multicolumn{2}{l}{~~~~~~~I''_s=\cup\{
          \shden{\methodsig{m}}(I'_s[\bar{v}/\inp{\methodsig{m}}])[\inp{\methodsig{m}}/\bar{v},\out/\res]
          ~|~ \methodsig{m} \mbox{ might be called here} \}
        }\\
        &\multicolumn{2}{l}{~~~~~~~I_1 = \{
          \REACHES{w_1}{w_2}~ | ~ (\REACHES{v_i}{v_j}{\in}
          I_m) \wedge (\PURE{v}_i {\in} I''_s)
          \wedge (\SHARE{w_1}{v_i} {\in} I'_s) \wedge} \\
        &\multicolumn{2}{l}{~~~~~~~~~~~~~~
          ((\REACHES{v_j}{w_2} {\in} I) \vee \ALIAS{w_2}{v_j} {\in}
          I'_s )\} 
        }
        \\
        &\multicolumn{2}{l}{~~~~~~~
          I_2 = \{ \REACHES{w_1}{w_2}~|~(\SHARE{v_i}{v_j}\in I'_s) \wedge 
          (\PURE{v}_i\in I''_s) \wedge
          (\SHARE{v_i}{w_1} \in I'_s) \wedge
          (\REACHES{v_j}{w_2}\in I) \}
        }\\
        &\multicolumn{2}{l}{~~~~~~~
          I_3 = \cup\{ (I_1{\cup}I_2)[v/\rho]~|~\ALIAS{v}{\rho} \mbox{
            after the call } \} }
        \\
        &\multicolumn{2}{l}{~~~~~~~
          I_4 = \{ \CYCLE{w} ~|~ (\SHARE{w}{v} \in I'_s) \wedge
          (\PURE{v}\in I''_s)\wedge (\CYCLE{v}\in I_m) \} }
        \\[4pt]
      \end{array} 
      \]
    \end{minipage}

    \caption{Abstract denotations for expressions}
  \label{fig:abs-sem-exp}
\end{figure}

\begin{figure}[t]
    \begin{minipage}{11.7cm}
\[
      \small
      \begin{array}{@{}l@{~}rl@{}}
        %
        %
        (1_c) & 
        \ASEMANTICS{\typenv}{\absinterp}{v\assign \mathit{exp}}(I) =& 
        (\exists v.\EXPASEMANTICS{\typenv}{\absinterp}{\mathit{exp}}(I))[\res/v] 
        \\
        %
        %
        (2_c) & 
        \ASEMANTICS{\typenv}{\absinterp}{v.f\assign \mathit{exp}}(I) =&
        \exists\res.(I'\cup I_r\cup I_c)~\mbox{where} \\
        & I'_0 = &
        \EXPASEMANTICS{\typenv}{\absinterp}{\mathit{exp}}(I)
        \\
        & I' = & \mathit{condRemove}(I'_0,v,f) \\
        &
          I_r = &\left\{ \REACHES{w_1}{w_2} ~|~
            ((\ALIAS{w_1}{v}{\in} I'_s) \vee (\REACHES{w_1}{v} {\in}
            I'))~\wedge \right.\\ & & 
            \left. ((\ALIAS{\res}{w_2}{\in} I'_s) \vee (\REACHES{\res}{w_2}
            {\in} I'))
          \right\}   \\
        & I_c =&\left\{ \CYCLE{w}~|~
            ((\REACHES{\res}{v} \in I') \vee (\ALIAS{\res}{v}{\in}
            I'_s) \vee
            (\CYCLE{\res} {\in} I'))~\wedge \right. \\ & & \left. 
            ((\ALIAS{w}{v}{\in} I'_s) \vee (\REACHES{w}{v} {\in} I'))
          \right\} 
        \\[0.5ex]
        %
        %
        (3_c) & \ASEMANTICS{\typenv}{\absinterp}{
          \begin{array}{@{}rl@{}}
            \ifte~\mathit{exp} & \iftethen~\mathit{com}_1 \\
            & \ifteelse~\mathit{com}_2 
          \end{array}
        }
        (I) = & 
        \ASEMANTICS{\typenv}{\absinterp}{\mathit{com}_1}(I) ~\cup~
        \ASEMANTICS{\typenv}{\absinterp}{\mathit{com}_2}(I) 
        \\
        %
        %
        (4_c) &
        \ASEMANTICS{\typenv}{\absinterp}{\while~\mathit{exp}~\whilebody~\mathit{com}}(I)
        =& \absden(I) \mbox{ where }
        \absden=\mathit{lfp}(\lambda w. \lambda
        I.w(\ASEMANTICS{\typenv}{\absinterp}{\mathit{com}}(I)))
        \\
        %
        %
        (5_c) & \ASEMANTICS{\typenv}{\absinterp}{\return~\mathit{exp}}(I) =& 
        \EXPASEMANTICS{\typenv}{\absinterp}{\mathit{exp}}(I)[\res/\out]
        \\
        %
        %
        (6_c) & \ASEMANTICS{\typenv}{\absinterp}{\mathit{com}_1; \mathit{com}_2}(I) =&
        \ASEMANTICS{\typenv}{\absinterp}{\mathit{com}_2}(\ASEMANTICS{\typenv}{\absinterp}{\mathit{com}_1}(I))
        \\
      \end{array} 
\]
    \end{minipage}

  \caption{Abstract denotations for commands}
  \label{fig:abs-sem-com}
\end{figure}

%% file: experiments.tex

\subsection{Note on an implementation}
\label{sec:experiments}

The present analysis has been implemented in the
COSTA~\cite{AlbertAGPZ07d} COSt and Termination Analyzer.  The
implementation works as a component of COSTA, and handles programs
written in full sequential \emph{Java bytecode}, which includes
control flow that originates from the handling of \emph{exceptions}.
\emph{Static fields} are accounted for as a kind of global variables:
this means that, for every class $\class$ and static field $f$, a
global variable $v_{\class.f}$ is added to the analysis (note that the
set of such global variables is statically decidable by simply
inspecting the program code).  The acyclicity information is used by
COSTA to prove the termination or infer the resource usage of
programs.

It is worth mentioning that the implementation is a prototype, and
that it can be optimized in many ways.  In fact, the present paper
focuses on the theoretical definition of an existing analysis, so that
the implementation is not the most important issue.
As a matter of fact, such implementation deals with a different
language with respect to the original implementation; this implies,
for example, having to account in a specific way for advanced features
of Java and Java bytecode like objects, exceptions, and static fields.
The single-field optimization discussed in Section
\ref{sec:fieldUpdate} is not implemented.

%% file: conclusions.tex
\section{Conclusions}
\label{sec:conclusions}

This paper discusses an acyclicity analysis of a Java-like language
with mutable data structures, based on reachability between variables.
In particular, the main focus of the paper is on the formalization of
an existing analysis within the framework of Abstract Interpretation.
The proposed acyclicity analysis is based on the observation that a
field update \lstinline!x.f=y! might create a new cycle iff \yy
reaches \xx or aliases with it before the command.
Two abstract domains are first defined, which capture the
\emph{may-reach} and \emph{may-be-cyclic} properties.  Then, an
abstract semantics which works on their reduced product is introduced:
it uses reachability information to improve the detection of
cyclicity, and cyclicity to improve the tracking of reachability.

The analysis is proven to be sound; i.e., no cyclic data structure are
ever considered acyclic.  It is also proven to be the best correct
approximation of the concrete semantics with respect to the chosen
abstraction.  Moreover, it can be shown to obtain precise results in a
number of non-trivial scenarios, where the sharing-based approach is
less precise~\cite{RossignoliS06}.  Indeed, since the existence of a
directed path between the locations bound to two variables implies
that such variables share, the proposed reachability-based analysis
will never be less precise than the sharing-based approach.  In
particular, it is worth noticing that the reachability-based approach
can often deal with directed acyclic graphs, whereas sharing-based
techniques will consider, in general, any DAG as cyclic.

%% file: proofs.tex

\section{Proofs}
\label{sec:proofs}

This appendix includes proofs for:
Lemma~\ref{lemma:gi-reach} in~\ref{sec:proofs:1};
Lemma~\ref{lemma:gi-cyc} in~\ref{sec:proofs:2};
Lemma~\ref{lemma:equivalenceClass} in~\ref{sec:proofs:3}; and
Theorem~\ref{th:soundness} in~\ref{sec:proofs:4}.

\subsection{Proof of Lemma~\ref{lemma:gi-reach}}
\label{sec:proofs:1}
\bigskip

\input proof1

\subsection{Proof of Lemma~\ref{lemma:gi-cyc}}
\label{sec:proofs:2}
\bigskip

\input proof2

\subsection{Proof of Lemma~\ref{lemma:equivalenceClass}}
\label{sec:proofs:3}
\bigskip

\input proof3

\subsection{Proof of Thoerem~\ref{th:soundness}}
\label{sec:proofs:4}
\bigskip

\input proof4

%

%% file: proof1.tex

  %
  Due to the definition of Galois insertion, the result to prove
  amounts to say that both
  \[ \begin{array}{r@{~~}l} (a) & \forall
    \abselemr\in\domr.~\alphar(\gammar(\abselemr)) = I_r \\
    \mbox{and}~(b) & \forall
    \concelem\in\condom.~\gammar(\alphar(\concelem)) \supseteq
    \concelem
  \end{array} \]
  hold, where $\subseteq$ is the ordering on $\condom$.
  
  \paragraph{Part (a)} We show that $\REACHES{v}{w} \in \abselemr
  \Leftrightarrow \REACHES{v}{w} \in \gammar(\alphar(\concelem))$.
  ($\Rightarrow$) assume $\REACHES{v}{w} \in \abselemr$; then,
  according to the definition of $\domr$ and class reachability, there
  must be a concrete state $\statesym\in\states{\typenv}$ in which $v$
  reaches $w$, since, otherwise, the statement $\REACHES{v}{w}$ cannot
  be part of the domain $\domr$.  We construct a state $\statesym'$
  from $\statesym$ by setting all reference variables but $v$ and $w$
  to \nil.
  By the definition of $\gammar$, this specific $\statesym'$ must be
  in $\gammar(\abselemr)$. This, according to the definition of
  $\alphar$, implies that
  $\REACHES{v}{w}\in \alphar(\gammar(\abselemr))$.
  ($\Leftarrow$) assume
  $\REACHES{v}{w} \in \alphar(\gammar(\abselemr))$.  According to the
  definition of $\alphar$, this means that there exists at least one
  $\statesym \in \gammar(\abselemr)$ in which $v$ reaches $w$, and,
  according to the definition of $\gammar$, this can only happen if
  $\REACHES{v}{w}\in\abselemr$.
 
  \paragraph{Part (b)} We show that
  $\statesym\in\concelem \Rightarrow \statesym\in\gammar(\alphar(\concelem))$.
  Let $\statesym \in \concelem$, and let $\abselemr$ be the set of all
  reachability relations in $\statesym$, i.e., $v$ reaches $w$ in
  $\statesym$ iff $\REACHES{v}{w}\in \abselemr$. Clearly,
  $\abselemr \subseteq \alphar(\concelem)$.  Then, according to the
  definition of $\gammar$, $\statesym$ must be in
  $\gammar(\alphar(\concelem))$ since it satisfies $\forall
  v,w\in\typenv.~
  v~\mbox{reaches}~w~\mbox{in}~\statesym \Rightarrow \REACHES{v}{w}\in \alphar(\concelem)$. \hfill \qed

%% file: proof2.tex

Very similar to the proof of Lemma~\ref{lemma:gi-reach}. \hfill\qed

%% file: proof3.tex

\paragraph{($\Rightarrow$)} We show that:
\[
\small
\begin{array}{@{}ccc@{}c@{}c@{}}
\gammarc(\tuple{\abselemr^1,\abselemc^1}){=}
    \gammarc(\tuple{\abselemr^2,\abselemc^2}) 
    &
    \Rightarrow 
    &
    \abselemc^1 {=} \abselemc^2  
    &
    \wedge  
    &
    (\abselemr^1
    {\setminus} \{ \REACHES{v}{v} \mid \CYCLE{v}{\notin} \abselemc^1 \}) {=}
    (\abselemr^2 {\setminus} \{ \REACHES{v}{v} \mid \CYCLE{v}{\notin}
    \abselemc^2 \})\\
 \underbrace{~~~~~~~~~~~~~~~~~~~~~~~~~~~~~~~~}
&&
 \underbrace{~~~~~~~~~}
&&
 \underbrace{~~~~~~~~~~~~~~~~~~~~~~~~~~~~~~~~~~~~~~~~~~~~~~~~~~~~~~~}\\
F && G && H\\
  \end{array} 
\]
  \noindent
  First, note that the logical formula $F \Rightarrow (G \wedge H)$ is
  equivalent to $(\lnot G \Rightarrow \lnot F) \wedge (\lnot
  H \Rightarrow \lnot F)$.  The proof is by contradiction, and
  consists of two parts:

  \begin{enumerate}
  \item proving that $I_c^1 \neq I_c^2$ implies
    $\gamma(\tuple{I_r^1,I_c^1}) \neq \gamma(\tuple{I_r^2,I_c^2})$; and
  \item proving that $(I_r^1 \setminus \{
    \REACHES{v}{v}|\CYCLE{v}{\notin}I_c^1 \}) \neq (I_r^2 \setminus \{
    \REACHES{v}{v}|\CYCLE{v}{\notin} I_c^2 \})$ implies\\
    $\gamma(\tuple{I_r^1,I_c^1}) \neq \gamma(\tuple{I_r^2,I_c^2})$.
  \end{enumerate}
  The proof goes as follows.
  \begin{enumerate}
    
  \item Suppose $I_c^1 \neq I_c^2$, and let $X_1=\{ v\;|\;\CYCLE{v}
    \in I_c^1 \setminus I_c^2\}$, and $X_2=\{ v\;|\;\CYCLE{v} \in
    I_c^2 \setminus I_c^1\}$.  Note that, by hypothesis, at least one
    of $X_1$ and $X_2$ must be non-empty.  For $i \in \{1,2\}$,
    let $\statesym_i$ be a state where
    \begin{itemize}
    \item[(a)] Every $v \in X_i$ is cyclic, but does not reach itself,
      and no other variable is cyclic; and
 
   \item[(b)] No variables reach any variables, i.e.,
     $\alphar(\{\statesym_i\})=\emptyset$.  Note that this requirement
     is consistent, since the cyclicity of some variables (in this
     case, those in $X_i$) does not necessarily imply the existence of
     a reachability path between variables.
   \end{itemize}
    
    \noindent It is easy to see that $\statesym_1$ and $\statesym_2$
    both belong to $\gamma_r(I_r^1) \cap \gamma_r(I_r^2)$, since they
    do not include any reachability statement; therefore, if
    $X_1 \neq \emptyset$, then $\statesym_1$ belongs to
    $\gamma(\tuple{I_r^1,I_c^1})$, but not to
    $\gamma(\tuple{I_r^2,I_c^2})$, since $\tuple{I_r^2,I_c^2}$ does
    not allow the cyclicity on variables from $X_1$.  Dually, if
    $X_2 \neq \emptyset$, then $\statesym_2$ belongs to
    $\gamma(\tuple{I_r^2,I_c^2})$ but not to
    $\gamma(\tuple{I_r^1,I_c^1})$.

  \item Suppose $R_1 =
    I_r^1 \setminus \{ \REACHES{v}{v} \mid \CYCLE{v}{\notin}I_c^1 \}$
    is different from $R_2 =
    I_r^2 \setminus \{ \REACHES{v}{v} \mid \CYCLE{v}{\notin}I_c^2 \}$,
    and let $S_1 = R_1 \setminus R_2$ and $S_2 = R_2 \setminus R_1$.
    Note that at least one between $S_1$ and $S_2$ is non-empty.  If
    $S_1$ is not empty, then let $p \in S_1$ be one of the statements
    which in $R_1$ but not in $R_2$.  A state $\statesym_1$ can be
    chosen such that
    \begin{itemize}
    \item[(a)] If $p = \REACHES{v}{v}$, then $v$ is the only cyclic
      variable in $\statesym_1$ (note that the cyclicity of $v$ must be
      allowed by $I_c^1$ since, otherwise, $p$ would not be included
      in $R_1$ and thus not in $S_1$ too); and
    \item[(b)] If $p=\REACHES{v}{w}$, with $v\neq w$; then, $v$ must
      reach $w$ in $\statesym_1$, and no other variable reaches any
      other variable.  Also, no variables can be cyclic.
    \end{itemize}
    Clearly, in both cases above such state belongs to
    $\gamma(\tuple{I_r^1,I_c^1})$, but it cannot be in
    $\gamma(\tuple{I_r^2,I_c^2})$ because: in (a), either
    $\REACHES{v}{v}\not\in I_r^2$ (so that
    $\sigma_1\not\in\gammar(I_r^2)$), or $\CYCLE{v}\not\in I_c^2$ (so
    that thus $\sigma_1\not\in\gammac(I_c^2)$); and, in (b),
    $\REACHES{v}{w}\not\in I_r^2$, so that
    $\sigma_1\not\in\gammar(I_r^2)$.
    Dually, if $S_2$ is empty, then $S_1$ cannot be empty, and, with a
    similar reasoning, a state $\statesym_2$ can be found which
    belongs to $\gamma(\tuple{I_r^2,I_c^2})$, but not to
    $\gamma(\tuple{I_r^1,I_c^1})$.  
  \end{enumerate}

\paragraph{($\Leftarrow$)} We prove that:
\[ 
\small
 \abselemc^1 {=} \abselemc^2  {\wedge}
    (\abselemr^1 {\setminus} \{ \REACHES{v}{v} | \CYCLE{v}{\notin}
    \abselemc^1 \}) {=} (\abselemr^2 {\setminus} \{
    \REACHES{v}{v} | \CYCLE{v}{\notin}
    \abselemc^2 \}) 
    \Rightarrow
    \gammarc(\tuple{\abselemr^1,\abselemc^1}){=}
    \gammarc(\tuple{\abselemr^2,\abselemc^2})
\]
\noindent 
It follows easily from observing that, under the hypothesis of the
above implication, the only difference between
$\tuple{\abselemr^1,\abselemc^1}$ and
$\tuple{\abselemr^2,\abselemc^2}$ is that
$\tuple{\abselemr^1,\abselemc^1}$ may contain some statements
$\REACHES{v}{v}$ for variables $v$ such that
$\CYCLE{v}\notin \abselemc^1$, and $\tuple{\abselemr^2,\abselemc^2}$
may contain some (different) statements $\REACHES{v}{v}$ for variables
$v$ such that or $\CYCLE{v}\notin \abselemc^2$.  However, adding such
statements to both abstract values does not change the set of concrete
states they represent, since the possibility that $v$ reaches itself
in any concrete state is contradicted by the lack of the $\CYCLE{v}$
statement.
In other words, there is no concrete state which belongs either to
$\gammarc(\tuple{\abselemr^1,\abselemc^1})$ or
$\gammarc(\tuple{\abselemr^2,\abselemc^2})$, but not to
$\gammarc(\tuple{\abselemr^1{\setminus} \{ \REACHES{v}{v}|\CYCLE{v}{\notin} \abselemc^1 \},~\abselemc^1})$
and
$\gammarc(\tuple{\abselemr^2{\setminus} \{ \REACHES{v}{v}|\CYCLE{v}{\notin} \abselemc^2 \},~\abselemc^2})$
(which are equal by the hypothesis.) \hfill\qed

%% file: proof4.tex

\newcommand{\gammarcrho}{\ensuremath{\gamma_{\mathit rc}^{\typenv\cup\{\res\}}}}
 
This proof of soundness amounts to proving the soundness of all
abstract denotations for expressions and commands, assuming that a
current interpretation $\interp$ and a corresponding abstract one
$\absinterp$ which correctly approximates $\interp$ are available.
Then, a simple induction can be applied to show that the abstract
semantics of Definition~\ref{def:conc-den-semantics} correctly
approximates the concrete semantics of
Definition~\ref{def:abs-den-semantics} (the induction step basically
applies the denotations on the elements of $\interp$ and
$\absinterp$).

In the following, let $\statesym$ be a concrete state, $\mathit{com}$
be a command, $\mathit{exp}$ be an expression, and $\statesym^*$ be
the state obtained by executing $com$ or evaluating $\mathit{exp}$ in
$\statesym$.
The soundness of the abstract denotations for expressions and commands
amounts to say that, if $I \in \domrc{\typenv}$ correctly approximates
$\statesym$, i.e., $\statesym \in \gammarc(I)$, then the abstract
state $I^* = \ASEMANTICS{\typenv}{\absinterp}{\mathit{com}}(I)$ (or
$I^* = \EXPASEMANTICS{\typenv}{\absinterp}{\mathit{exp}}(I)$, in the
case of expressions) correctly approximates $\statesym^*$.
Formally, we show that

\begin{enumerate}
\item $\forall \statesym\in\states{\tau}, I\in\domrc{\typenv}.~
  \statesym \in
  \gammarcrho(I)~\Rightarrow~\einter{\typenv}{\iota}{\mathit{exp}}(\statesym)
  \in
  \gammarcrho(\EXPASEMANTICS{\typenv}{\absinterp}{\mathit{exp}}(I))$
\item $\forall \statesym\in\states{\tau}, I\in\domrc{\typenv}.~
  \statesym \in
  \gammarc(I)~\Rightarrow~\cinter{\typenv}{\iota}{\mathit{com}}(\statesym)
  \in \gammarc(\ASEMANTICS{\typenv}{\absinterp}{\mathit{com}}(I))$
\end{enumerate}
  
\noindent
Note that, if $\statesym^*$ is obtained after evaluating an
expression, then $\res\in\dom(\statesym^*)$, while, if it is obtained
after executing a command, then $\dom(\statesym^*)=\dom(\statesym)$.

The soundness proof considers separately the rules of the abstract
semantics $\EXPASEMANTICS{\typenv}{\absinterp}{\_}(\_)$ and
$\ASEMANTICS{\typenv}{\absinterp}{\_}(\_)$.  When some logical fact is
said to hold \emph{by soundness}, it means that it holds by the
hypothesis on the input (i.e., that $\statesym \in \gammarc(I)$
holds), or by induction on sub-expressions or sub-commands.  For
example, the fact that $v$ reaches $w$ in $\statesym$ implies
$\REACHES{v}{w} \in I$ \emph{by soundness}, since $I$ is supposed to
be a sound description of $\statesym$.

\paragraph{Denotations $(1_e)$, $(2_e)$, and $(3_e)$} Suppose
  $\statesym^*\not\in \gammarcrho(I^*)$. Then, according to the
definition of $\gammarcrho$, it must be the case that (i) $w_1$
reaches $w_2$ in $\statesym^*$ but $\REACHES{w_1}{w_2} \not\in I^*$;
or (ii) $w$ is cyclic in $\statesym^*$ but $\CYCLE{w}\not\in I^*$.
This contradicts the soundness hypothesis $\statesym \in \gammarc(I)$,
since $I^*=I$ and $\statesym$ and $\statesym^*$ have the same
reachability and cyclicity information\footnote{Note, that, unlike in
  Java, the simple act of creating an object does not involve, in
  itself, any action on its content, i.e., there are no side effects
  due to the constructor.}.
\paragraph{Denotation $(4_e)$} Assume $\tau(v)\neq\integer$,
otherwise the reasoning we developed for case $(1_e)$ applies.  Note
that this case does not have any side effects, except defining the new
variable $\res$.
If $\sigma^*\not\in \gammarcrho(I^*)$, then, according to the
definition of $\gammarcrho$, it must be the case that (i) $w_1$
reaches $w_2$ in $\statesym^*$ but $\REACHES{w_1}{w_2} \not\in I^*$;
or (ii) $w$ is cyclic in $\statesym^*$ but $\CYCLE{w}\not\in I^*$.
Suppose we are in case (i):

\begin{itemize}

\item If $w_1\neq\res\wedge w_2\neq\res$, then
  $\statef{\statesym}(w_2)=\statef{\statesym}^*(w_2)\in
  \reachable{\statef{\statesym}^*(w_1)}{\statem{\statesym}^*}=
  \reachable{\statef{\statesym}(w_1)}{\statem{\statesym}}$, i.e.,
  $w_1$ reaches $w_2$ in $\statesym$.  By the soundness hypothesis
  $\statesym \in \gammarc(I)$ we have $\REACHES{w_1}{w_2}\in I
  \subseteq I^*$, which contradicts $\REACHES{w_1}{w_2} \not\in I^*$.

\item If $w_1=\res \wedge w_2\neq\res$, then
  $\statef{\statesym}(w_2)=\statef{\statesym}^*(w_2)\in
  \reachable{\statef{\statesym}^*(\res)}{\statem{\statesym}^*}=
  \reachable{\statef{\statesym}(v)}{\statem{\statesym}}$, i.e., $v$
  reaches $w_2$ in $\statesym$.
  By the soundness hypothesis $\statesym \in \gammarc(I)$, we have
  $\REACHES{v}{w_2}\in I$ and thus $\REACHES{\res}{w_2}\in
  I[v/\res]\subseteq I^*$, which contradicts $\REACHES{\res}{w_2}
  \not\in I^*$.

\item If $w_1\neq\res \wedge w_2=\res$, then
  $\statef{\statesym}(v)=\statef{\statesym}^*(\res)\in
  \reachable{\statef{\statesym}^*(w_1)}{\statem{\statesym}^*}=
  \reachable{\statef{\statesym}(w_1)}{\statem{\statesym}}$, i.e.,
  $w_1$ reaches $v$ in $\statesym$.
  By the soundness hypothesis $\statesym \in \gammarc(I)$, we have
  $\REACHES{w_1}{v}\in I$ and thus $\REACHES{w_1}{\res}\in
  I[v/\res]\subseteq I^*$, which contradicts $\REACHES{w_1}{\res}
  \not\in I^*$.

\item If $w_1=\res \wedge w_2=\res$, then
  $\statef{\statesym}(v)=\statef{\statesym}^*(\res)\in
  \reachable{\statef{\statesym}^*(\res)}{\statem{\statesym}^*}=
  \reachable{\statef{\statesym}(v)}{\statem{\statesym}}$, i.e., $v$
  reaches $v$ in $\statesym$.
  By the soundness hypothesis $\statesym \in \gammarc(I)$, we have
  $\REACHES{v}{v}\in I$ and thus $\REACHES{\res}{\res}\in
  I[v/\res]\subseteq I^*$, which contradicts $\REACHES{\res}{\res}
  \not\in I^*$.

\end{itemize}
For case (ii), the reasoning is basically as (i), by considering
cyclicity instead of reachability.

\paragraph{Denotation $(5_e)$} Assume $f$ is of reference type,
otherwise the reasoning we have done for case $(1_e)$ applies.
Note that this case does not have any side effects, except defining
the new variable $\res$.
If $\sigma^*\not\in \gammarcrho(I^*)$, then, according to the
definition of $\gammarcrho$, it must be the case that (i) $w_1$
reaches $w_2$ in $\statesym^*$ but $\REACHES{w_1}{w_2} \not\in I^*$;
or (ii) $w$ is cyclic in $\statesym^*$ but $\CYCLE{w}\not\in I^*$.
Suppose we are in case (i):

\begin{itemize}

\item If $w_1\neq\res\wedge w_2\neq\res$, then
  $\statef{\statesym}(w_2)=\statef{\statesym}^*(w_2)\in
  \reachable{\statef{\statesym}^*(w_1)}{\statem{\statesym}^*}=
  \reachable{\statef{\statesym}(w_1)}{\statem{\statesym}}$, i.e.,
  $w_1$ reaches $w_2$ in $\statesym$.  By the soundness hypothesis
  $\statesym \in \gammarc(I)$, we have $\REACHES{w_1}{w_2}\in I
  \subseteq I^*$, which contradicts $\REACHES{w_1}{w_2} \not\in I^*$.
  
\item If $w_1=\res \wedge w_2\neq\res$, then
  $\statef{\statesym}(w_2)=\statef{\statesym}^*(w_2)\in
  \reachable{\statef{\statesym}^*(\res)}{\statem{\statesym}^*}
  \subseteq \reachable{\statef{\statesym}(v)}{\statem{\statesym}}$,
  i.e., $v$ reaches $w_2$ in $\statesym$.
  By the soundness hypothesis $\statesym \in \gammarc(I)$, we have
  $\REACHES{v}{w_2}\in I$ and thus $\REACHES{\res}{w_2}\in
  I[v/\res]\subseteq I^*$, which contradicts $\REACHES{\res}{w_2}
  \not\in I^*$.

\item If $w_1\neq\res \wedge w_2=\res$, then
  $\statef{\statesym}^*(\res)\in
  \reachable{\statef{\statesym}(w_1)}{\statem{\statesym}^*}=\reachable{\statef{\statesym}(w_1)}{\statem{\statesym}}$,
  we also have $\statef{\statesym}^*(\res)\in
  \reachable{\statef{\statesym}(v)}{\statem{\statesym}} )$ (since
  $\res=v.f$), i.e., $w_1$ shares with $v$ in $\statesym$. Thus,
  $\REACHES{w_1}{\rho}\in\{ \REACHES{w}{\rho} \mid \SHARE{w}{v} \in
  I_s\} \subseteq I^*$, which contradicts $\REACHES{w_1}{\rho} \not\in
  I^*$.

\item If $w_1=\res \wedge w_2=\res$, then
  $\statef{\statesym}^*(\res)\in
  \reachable{\statef{\statesym}^*(\res)}{\statem{\statesym}} )$,
  which means that $v$ is cyclic in $\statesym$, and by the
  soundness hypothesis we have $\CYCLE{v}\in I$, and thus
  $\REACHES{\res}{\res}\in \{ \REACHES{\res}{\res} | \CYCLE{v}\in I
  \}\subseteq I^*$, which contradicts $\REACHES{w_1}{\rho} \not\in
  I^*$.

\end{itemize}
For case (ii), the reasoning is basically as (i), by considering
cyclicity instead of reachability.

\paragraph{Denotation  $(6_e)$}
The proof for this case is by structural induction on expressions,
where the base-case include the non-compound expressions of cases
$(1_e)$-$(5_e)$ and $(7_e)$, for which we have seen already (case
$(7_e)$ is done below) that the abstract denotations correctly
approximate the concrete ones.
Let $I_1=\EXPASEMANTICS{\typenv}{\absinterp}{\mathit{exp_1}}(I)$ and
$\statesym_1=\einter{\typenv}{\iota}{\mathit{exp_1}}(\statesym)$.  By
the (structural) induction hypothesis, we have $\statesym_1 \in
\gammarcrho(I_1)$.  Moreover, since the state
$\state{\statef{\statesym}}{\statem{\statesym}_1}$ is basically
obtained by removing $\rho$ from $\statesym_1$, we also have
$\state{\statef{\statesym}}{\statem{\statesym}_1}\in
\gammarc(\exists\res.I_1)$.
Now, let
$I_2=\EXPASEMANTICS{\typenv}{\absinterp}{\mathit{exp_2}}(\exists\res.I_1)$,
and
$\statesym_2=\einter{\typenv}{\iota}{\mathit{exp_2}}(\state{\statef{\statesym}}{\statem{\statesym}_1})$;
then, by the (structural) induction hypothesis, we have $\statesym_2
\in \gammarcrho(I_2)$.
Since $\statesym^*$ is obtained from $\statesym_2$ by setting $\res$
to a number (i.e., there is no reachability or cyclicity relations in
$\statesym^*$ that involve $\res$), and since $I^*=\exists\res.I_2$,
we can conclude that $\statesym^* \in \gammarcrho(I^*)$.

\paragraph{Denotation $(7_e)$}
Calling a method $\methodsig{m}$ consists of an abstract execution of
its body on the actual parameters, followed by the propagation of the
effects of $\methodsig{m}$ to the calling context (i.e., the input
abstract state $I$).
First, note that, in the abstract semantics, reachability and
cyclicity statements are only removed when a variable is assigned.
Due to the use of shallow variables for the parameters, statements
about the formal parameters of $\methodsig{m}$ are never removed
during an abstract execution of its body.
Therefore, if, during the execution of $\methodsig{m}$, the variable
$v$ reaches $w$, then, at the end of the method, $v$ will be said to
possibly reach $w$, even if this reachability is destroyed at some
subsequent program point.  This is similar to the way sharing
information is dealt with in the present approach
(following~\cite{spoto:pair_sharing,GenaimS08}).
   
Keeping track of cyclicity is rather easy.  In addition to keeping all
cyclicity which is in $I$, a safe approximation is taken, which states
that, if an argument $v$ might become cyclic during the execution of
$\methodsig{m}$, then anything that shares with it before the
execution might also become cyclic.  This is accounted for in the
definition of $I_4$, and is clearly safe.  In fact, variables of the
calling method which are not arguments of the call, and do not share
with any argument $v_i$, cannot be affected by the execution of
$\methodsig{m}$.
  
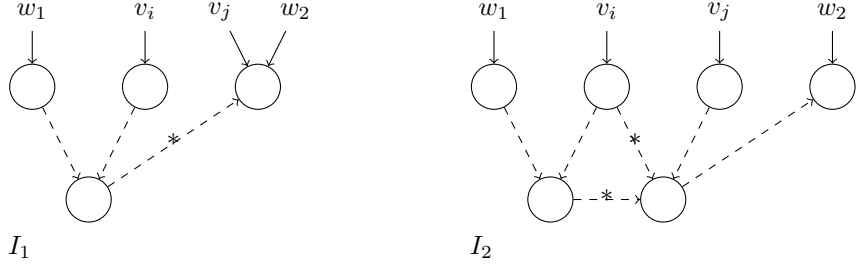
\begin{figure}
  \begin{minipage}{0.5\textwidth}
    \begin{tikzpicture}
      \tikzstyle{mylocation}=[draw,circle,minimum size=6mm]
      \tikzstyle{myvariable}=[minimum size=5mm]
      \tikzstyle{myreachability}=[->,dashed]
      \tikzstyle{myderef}=[->,double]
      
      \node[mylocation] (w1loc) at (0,0) {};
      \node[mylocation] (viloc) at (1.5,0) {};
      \node[mylocation] (vjw2loc) at (3,0) {};
      
      \node[myvariable] (w1) at (0,1) {$w_1$};
      \node[myvariable] (vi) at (1.5,1) {$v_i$};
      \node[myvariable] (vj) at (2.5,1) {$v_j$};
      \node[myvariable] (w2) at (3.5,1) {$w_2$};
      
      \node[mylocation] (w1viloc) at (0.75,-1.5) {};
      
      \draw[->] (w1) -- (w1loc);        
      \draw[->] (vi) -- (viloc);        
      \draw[->] (vj) -- (vjw2loc);        
      \draw[->] (w2) -- (vjw2loc);        
      
      \draw[myreachability] (w1loc) -- (w1viloc);
      \draw[myreachability] (viloc) -- (w1viloc);
      \draw[myreachability] (w1viloc) -- node {*} (vjw2loc);
    \end{tikzpicture}
    
    $I_1$
  \end{minipage}
  \begin{minipage}{0.5\textwidth}
    \begin{tikzpicture}
      \tikzstyle{mylocation}=[draw,circle,minimum size=6mm]
      \tikzstyle{myvariable}=[minimum size=5mm]
      \tikzstyle{myreachability}=[->,dashed]
      \tikzstyle{myderef}=[->,double]
      
      \node[mylocation] (w1loc) at (0,0) {};
      \node[mylocation] (viloc) at (1.5,0) {};
      \node[mylocation] (vjloc) at (3,0) {};
      \node[mylocation] (w2loc) at (4.5,0) {};
      
      \node[myvariable] (w1) at (0,1) {$w_1$};
      \node[myvariable] (vi) at (1.5,1) {$v_i$};
      \node[myvariable] (vj) at (3,1) {$v_j$};
      \node[myvariable] (w2) at (4.5,1) {$w_2$};
      
      \node[mylocation] (w1viloc) at (0.75,-1.5) {};
      \node[mylocation] (vivjloc) at (2.25,-1.5) {};
      
      \draw[->] (w1) -- (w1loc);        
      \draw[->] (vi) -- (viloc);        
      \draw[->] (vj) -- (vjloc);        
      \draw[->] (w2) -- (w2loc);        
      
      \draw[myreachability] (w1loc) -- (w1viloc);
      \draw[myreachability] (viloc) -- (w1viloc);
      \draw[myreachability] (viloc) -- node {*} (vivjloc);
      \draw[myreachability] (vjloc) -- (vivjloc);
      \draw[myreachability] (vivjloc) -- (w2loc);
      
      \draw[myreachability] (w1viloc) -- node {*} (vivjloc);
    \end{tikzpicture}
    
    $I_2$
  \end{minipage}
  
  \caption{Scenarios where a path from $w_1$ to $w_2$ can be created
    inside $\methodsig{m}$.  Dashed arrows represent reachability:
    they connect a variable to a reachable location (represented as a
    circle).  Solid arrows connect a variable $u$ to the location
    $\statef{\statesym}(u)$ directly bound to it.  Arrows labeled with
    * are paths which are created inside $\methodsig{m}$ (strictly
    speaking, they could also exist before the call), while the others
    existed before the method call.  In both cases, it can be seen
    that a reachability path from $w_1$ to $w_2$ is created, which
    contains a sub-path created inside $\methodsig{m}$ by modifying
    its arguments.}
  \label{fig:createPath}
\end{figure}

The treatment of reachability is more complicated: in addition to $I$
and $I_m$ (which is introduced by the method for $\bar{v}$), it is
necessary to take into account the effect of the method call on
variables which are not arguments.  This is done in the definition of
$I_1$, $I_2$, and $I_3$, which model the effects of $\methodsig{m}$ on
variables which share with its actual arguments.
Consider two arguments $v_i$ and $v_j$ (where $i$ can be equal to
$j$): a path between two variables $w_1$ and $w_2$ (which can be
arguments, or non-argument variables) can be created by
$\methodsig{m}$ if (i) $v_i$ and $w_1$ share before the call, $v_j$
and $w_2$ alias before the call, $v_i$ is modified in $\methodsig{m}$,
and $v_i$ reaches $v_j$ after the call; or (ii) $v_i$ and $w_1$ share
before the call, $v_j$ reaches $w_2$ before the call, $v_i$ is
modified in $\methodsig{m}$, and $v_i$ and $v_j$ share (without
reaching each other) after the call.  The two cases are accounted for
in the definition of, resp., $I_1$ and $I_2$, and are depicted in
Fig.\ref{fig:createPath}.
In both cases, the creation of the path requires that an argument is
modified in $\methodsig{m}$ (condition $\PURE{v}_i\in sh'$), and that
$v_i$ and $v_j$ do not point to disjoint regions of the heap (i.e.,
either $v_i$ reaches $v_j$, or they simply share).  As a result, if
these conditions are met, then the statement $\REACHES{w_1}{w_2}$ is
added.
It can be seen that this accounts for all cases where some change in
the arguments of $\methodsig{m}$ affects the reachability between
non-argument variables.
%
%

\noindent
Finally, $I_3$ considers all variables $v$ aliasing with the return
value at the end of $\methodsig{m}$ (note that these are the only new
aliasing statements involving arguments which can be created in the
body of $\methodsig{m}$) : the information about them is cloned for
$\res$.

\paragraph{Denotation $(1_c)$}
Suppose $\statesym^*\not\in \gammarcrho(I^*)$. Then, according to the
definition of $\gammarcrho$, it must be the case that (i) $w_1$ reaches
$w_2$ in $\statesym^*$ but $\REACHES{w_1}{w_2} \not\in I^*$; or (ii)
$w$ is cyclic in $\statesym^*$ but $\CYCLE{w}\not\in I^*$.
Suppose we are in case (i), and let
$\statesym_e=\einter{\typenv}{\iota}{\mathit{exp}}(\statesym)$ and
$I_1=\EXPASEMANTICS{\typenv}{\absinterp}{\mathit{exp}}(I)$.

\begin{itemize}
  
\item If $w_1\neq v \wedge w_2\neq v$, then it must be the case that
  $w_1$ reaches $w_2$ in $\statesym_e$.  By the soundness of the
  expressions denotations we must have $\statesym_e \in
  \gammarcrho(I_1)$, which means that $\REACHES{w_1}{w_2}\in I_1$;
  thus, $\REACHES{w_1}{w_2}\in (\exists v.I_1)[\res/v]=I^*$, which
  contradicts $\REACHES{w_1}{w_2}\not\in I^*$.

\item If $w_1=v \wedge w_2\neq v$, then it must be the case that
  $\res$ reaches $w_2$ in $\statesym_e$.  By the soundness of the
  denotations for expressions, we must have $\statesym_e \in
  \gammarcrho(I_1)$, which means that $\REACHES{\res}{w_2}\in I_1$,
  and thus $\REACHES{v}{w_2}\in (\exists v. I_1)[\res/v]=I^*$, which
  contradicts $\REACHES{v}{w_2}\not\in I^*$.

\item If $w_1\neq v \wedge w_2= v$, then it must be the case that
  $w_1$ reaches $\res$ in $\statesym_e$.  By the soundness of the
  denotations for expressions, we must have $\statesym_e \in
  \gammarcrho(I_1)$, which means that $\REACHES{w_1}{\res}\in I_1$,
  and thus $\REACHES{w_1}{v}\in (\exists v. I_1)[\res/v]=I^*$, which
  contradicts $\REACHES{w_1}{v}\not\in I^*$.
  
\item If $w_1=v \wedge w_2=v$, then it must be the case that $\res$
  reaches $\res$ in $\statesym_e$.  By the soundness of the
  denotations for expressions, we must have $\statesym_e \in
  \gammarcrho(I_1)$, which means that $\REACHES{\res}{\res}\in I_1$,
  and thus $\REACHES{v}{v}\in (\exists v. I_1)[\res/v]=I^*$, which
  contradicts $\REACHES{v}{v}\not\in I^*$.

\end{itemize}
Case (ii) can be done with similar reasoning.

\begin{figure}
  \begin{minipage}{58mm}
    \begin{tikzpicture}
      \tikzstyle{mylocation}=[draw,circle,minimum size=6mm]
      \tikzstyle{myvariable}=[minimum size=5mm]
      \tikzstyle{myreachability}=[->,dashed]
      \tikzstyle{myderef}=[->,double]
      
      \node[mylocation] (loc) at (0,0) {$\ell_e$};
      \node[myvariable] (v) at (-1,1) {$v$};
      \node[myvariable] (u1) at (0,-1) {$u_1$};
      \node[myvariable] (u2) at (-2,0) {$u_2$};
      \node[mylocation] (loc2) at (2,0) {$\ell_3$};
      \node[myvariable] (u3) at (2,1) {$u_3$};
      \node[mylocation] (loc3) at (1.5,-1.5) {$\ell_4$};
      \node[myvariable] (u4) at (3.5,-1.5) {$u_4$};
      
      \draw[->] (v) -- (loc);
      \draw[->] (u1) -- (loc);
      \draw[myreachability] (u2) -- (loc);
      \draw[myreachability] (loc) -- (loc2);
      \draw[myreachability] (loc) -- (loc3);
      \draw[->] (u3) -- (loc2);
      \draw[myreachability] (u4) -- (loc3);
      \path (loc) edge [myderef,loop above] node {$f$} ();
    \end{tikzpicture}
  \end{minipage}
  \begin{minipage}{63mm}
    \begin{tikzpicture}
      \tikzstyle{mylocation}=[draw,circle,minimum size=6mm]
      \tikzstyle{myvariable}=[minimum size=5mm]
      \tikzstyle{myreachability}=[->,dashed]
      \tikzstyle{myderef}=[->,double]
      
      \node[mylocation] (vloc) at (0,0) {};
      \node[mylocation] (loc) at (1.5,0) {$\ell_e$};
      \node[myvariable] (v) at (-1,1) {$v$};
      \node[myvariable] (u1) at (-1,-1) {$u_1$};
      \node[myvariable] (u2) at (-2,0) {$u_2$};
      \node[myvariable] (u3) at (1,-1) {$u_3$};
      \node[myvariable] (u4) at (2,-2) {$u_4$};
      \node[myvariable] (u5) at (0.5,1.5) {$u_5$};
      \node[mylocation] (loc2) at (0,-2) {$\ell$};
      \node[mylocation] (loc3) at (1.5,2) {$\ell$};
      \node[myvariable] (u6) at (2.5,-1) {$u_6$};
      \node[myvariable] (u7) at (3.5,0) {$u_7$};
      \node[myvariable] (u8) at (3.5,2) {$u_8$};
      
      \draw[->] (v) -- (vloc);
      \draw[myderef] (vloc) -- node[auto] {$f$} (loc);
      \draw[->] (u1) -- (vloc);
      \draw[->] (u6) -- (loc);
      \draw[myreachability] (u2) -- (vloc);
      \draw[myreachability] (loc) -- (u7);
      \draw[myreachability] (vloc) -- (loc2);
      \draw[myreachability] (loc) -- (loc3);
      \draw[myreachability] (u5) -- (loc);
      \draw[myreachability] (u4) -- (loc2);
      \draw[myreachability] (u8) -- (loc3);
      \draw[myreachability] (vloc) -- (u3);
    \end{tikzpicture}
  \end{minipage}
  
  \caption{The possible scenarios for case $(2_c)$: (a) $\ell_e$ and
    $\statef{\statesym}(v)$ coincide (left); and (b) they do not
    coincide.  Variables $u_i$ represent the possible relations
    between the variable $u$ used in the proof and the data structure
    modified by the field update.  Double solid arrows stand for field
    dereferencing, and are labeled with the name of the field.  For
    the other kinds of arrows, see Fig.~\ref{fig:createPath}. }
    \label{fig:putfieldScenarios}
\end{figure}
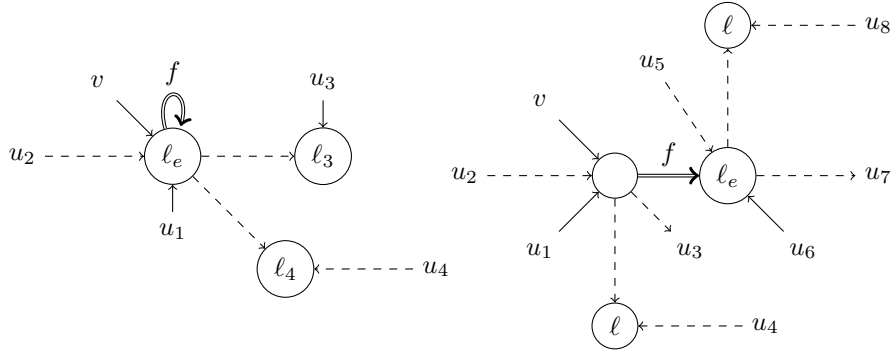

\paragraph{Denotation $(2_c)$}
This case is trivial when $f$ has type \integer, since only side
effects during the evaluation of $\mathit{exp}$ have to be taken into
account.  If $f$ has reference type, then this command is equivalent
to first evaluating $\mathit{exp}$, and then executing $v.f:=\res$.
Let $\statesym' = \einter{\typenv}{\interp}{\mathit{exp}}(\statesym)$,
and $\ell_e=\statef{\statesym}'(\res)$.
If $v$ and $\ell_e$ are considered, then there are two main cases
(Fig.~\ref{fig:putfieldScenarios}): (a) $\statef{\statesym}(v) =
\ell_e$; or (b) $\statef{\statesym}(v) \neq \ell_e$.

\begin{itemize}
  
\item[(a)] In this case, a cycle on $v$ is created, whose length is 1.
  If another variable $u$ (possibly, $v$ itself) sharing with $v$ in
  $\statesym^*$ is considered, then there are several possible
  scenarios in the heap, and soundness has to be proven for each of
  them.
  
  \begin{itemize}
    
  \item $u$ aliases with $v$ or reaches $v$ (cases $u_1$ and $u_2$ in
    the left-hand side of Fig.~\ref{fig:putfieldScenarios}).  In this
    case, $u$ reaches $v$ via $f$, and this is taken into account in
    the definition of $I_r$, where $u$ plays the role of $w_1$, and
    $v$ also plays the role of $w_2$.  The result is that $I_r$
    includes $\REACHES{u}{v}$, as expected.  The semantics correctly
    adds $\REACHES{v}{v}$ as well (in fact, $v$ can play the role of
    both $w_1$ and $w_2$).  As for cyclicity, the definition of $I_c$
    guarantees that $\CYCLE{v}$ and $\CYCLE{u}$ will belong to $I^*$.

  \item $v$ reaches $u$ (case $u_3$ in the same figure).  In this
    case, $\REACHES{v}{u} \in I^*$ since, in the definition of $I_r$,
    $u$ plays the role of $w_2$ (note that $v$ and $\res$ alias).  $v$
    will also be considered as cyclic by the definition of $I_c$;

  \item $v$ and $u$ both reach a common location $\ell$ (case $u_4$).
    If none of the previous cases happens, then $v$ and $u$ do not
    reach each other, so that $I^*$ does not need to contain
    reachability statements between them.  In general, only $v$ will
    be considered as cyclic in this case (in the same way as the
    previous cases).

  \end{itemize}
  
\item[(b)] In this case, when considering $u$, the number of possible
  scenarios for reachability is larger.  Moreover, there are two
  scenarios where $v$ would be cyclic \emph{after} the update
  (i) $\ell_e$ reaches $v$, so that a cycle is created by the field
  update, and $v$ becomes cyclic (if it was not already); or (ii)
  $\ell_e$ does not reach $v$, so that $v$ is cyclic only if it was
  already cyclic in $\statesym$, and the same applies to $\ell_e$.  In
  case (ii), it can be easily seen that the definition of $I_c$
  accounts for the cyclicity of $v$ since $\CYCLE{v}$ belongs to $I$
  by soundness and will not be removed.  Case (i) will be discussed in
  the following, for each scenario.

  \begin{itemize}
    
  \item $u$ reaches $v$ or aliases with it (cases $u_1$ and $u_2$ in
    the right-hand side of Fig.~\ref{fig:putfieldScenarios}).  In this
    case, it was also reaching $v$ (or aliasing with it) in
    $\statesym'$, so that (in the case of reachability)
    $\REACHES{u}{v} \in I'$, which implies $\REACHES{u}{v} \in I''$,
    as soundness requires.  As for cyclicity, in case (i), the
    cyclicity of $u$ is detected because it reaches $v$.
      
  \item Cases $u_3$, $u_4$, and $u_5$.  These cases are easy, because
    nothing changes with respect to the reachability between $u$ and
    $v$, and all the statements were already contained in $I$.
      
  \item $u$ points to $\ell_e$ or is reached by it (cases $u_6$ and
    $u_7$).  In this case, $u$ plays the role of $w_2$ in the
    definition of $I_r$, and is correctly considered to be reached by
    $v$.  As for cyclicity, $u$ will only become cyclic in case (i) if
    it points to $\ell_e$, or belongs to the cyclic path.  In both
    cases, the semantics accounts for it since $u$ would reach $v$,
    thus being considered as cyclic (definition of $I_c$).
      
  \item $w$ and $\ell_e$ reach some common location $\ell$ (case
    $u_8$).  Also easy since nothing changes with respect to the
    reachability between $u$ and $v$.
      
  \end{itemize}
\end{itemize}
  
\noindent 
Note that, due to the discussion in Section \ref{sec:fieldUpdate}, the
single-field optimization introduced by $\mathit{condRemove}$ is not
problematic for soundness, since the removal of statements is only
applied if the required conditions about $v$ and $f$ are guaranteed to
hold.  In any case, the conservative choice of taking
$\mathit{condRemove}(I'_0,v,f)$ to be $I'_0$ itself is also sound.

\paragraph{Denotation $(3_c)$} This case is quite straightforward,
given the inductive hypothesis on $\mathit{com}_1$ and
$\mathit{com}_2$, and the assumption that $\mathit{exp}$ has no side
effects and returns an \integer.
Suppose
$\statesym^*=\cinter{\typenv}{\iota}{\mathit{com_i}}(\statesym)$ for
$i\in\{1,2\}$, then, by the induction hypothesis, $\statesym^* \in
\gammarc( \ASEMANTICS{\typenv}{\absinterp}{\mathit{com_i}}(I))
\subseteq \gammarc(
\ASEMANTICS{\typenv}{\absinterp}{\mathit{com_i}}(I))\cup \gammarc(
\ASEMANTICS{\typenv}{\absinterp}{\mathit{com_2}}(I))=I^*$.
   
\paragraph{Denotations $(4_c)$, $(5_c)$, and $(6_c)$} Rules for
loops and concatenation are easy, given the inductive hypothesis on
the sub-commands, and the definition of the fixpoint. The rule for the
\return command is also easy, being basically similar to variable
assignment.

Having proven that all abstract denotations are sound with respect to
the concrete denotational semantics, together with Definition
\ref{def:abs-den-semantics} and the definition of a denotational
semantics, proves the theorem.  \hfill\qed
  %